\documentclass[preprint,12pt]{elsarticle}

\usepackage[utf8]{inputenc}
\usepackage{amsmath,amsfonts,amsthm} 
\usepackage{amssymb}
\usepackage{graphicx}
\usepackage{xfrac}
\usepackage{color}
\usepackage{comment}
\usepackage{appendix}

\usepackage{subcaption}
\usepackage{amsthm}
\newcommand{\half}{{\tfrac{1}{2}}}
\newcommand{\threehalf}{{\tfrac{3}{2}}}

\newcommand{\K}{\mathcal{K}}
\newcommand{\bx}{\mathbf{x}}
\newcommand{\bn}{\mathbf{n}}
\newcommand{\bu}{\mathbf{u}}
\newcommand{\bv}{\mathbf{v}}
\newcommand{\bXX}{\mathbf{X}}
\newcommand{\bUU}{\mathbf{U}}
\newcommand{\p}{\partial}
\newcommand{\be}{\begin{equation}}
\newcommand{\ee}{\end{equation}}
\newcommand{\Ord}{O}

\newcommand{\inblue}[1]{{\color{blue}{#1}}}

\newtheorem*{remark}{Remark}

\def\Xint#1{\mathchoice
   {\XXint\displaystyle\textstyle{#1}}%
   {\XXint\textstyle\scriptstyle{#1}}%
   {\XXint\scriptstyle\scriptscriptstyle{#1}}%
   {\XXint\scriptscriptstyle\scriptscriptstyle{#1}}%
   \!\int}
\def\XXint#1#2#3{{\setbox0=\hbox{$#1{#2#3}{\int}$}
     \vcenter{\hbox{$#2#3$}}\kern-.5\wd0}}

\def\dashint{\Xint-}
\usepackage{graphicx}
\usepackage{float}
\usepackage{geometry}
\journal{xxx}

\begin{document}

\begin{frontmatter}

\title{A fast mesh-free boundary integral method for two-phase flow with soluble surfactant}

\author[njit]{Samantha G. Evans}
\ead{sge7@njit.edu}
\author[njit]{Michael Siegel\corref{cor1}}
 \ead{michael.s.siegel@njit.edu}
\cortext[cor1]{Corresponding author}
\author[smu]{Johannes Tausch}
 \ead{tausch@mail.smu.edu}
\author[njit]{Michael R. Booty}
 \ead{booty@njit.edu}

\affiliation[njit]{organization={Department of Mathematical Sciences and Center for Applied Mathematics and
Statistics, New Jersey Institute of Technology},
            addressline={University Heights}, 
            city={Newark},
            state={NJ},
            postcode={07102}, 
            country={USA}}

\affiliation[smu]{organization={Department of Mathematics, Southern Methodist University},
            addressline={209 A Clements Hall}, 
            city={Dallas},
            state={TX},
            postcode={75275}, 
            country={USA}}

\begin{abstract}
    
We present an accurate and efficient boundary integral (BI) method for simulating the deformation of drops and bubbles in Stokes flow with soluble surfactant. 
Soluble surfactant advects and diffuses in bulk fluids while adsorbing and desorbing at interfaces.  Since the fluid velocity is coupled to the surfactant concentration, the advection-diffusion equation governing the bulk surfactant concentration $C$ is nonlinear, precluding the Green's function formulation necessary for a BI method. 
However, in the physically representative large P\'eclet number limit, an analytical reduction of the surfactant dynamics  permits a Green's function formulation for $C$
as an Abel-type time-convolution integral at each Lagrangian interface point.  A challenge in developing a practical numerical method based on this formulation is the fast evaluation of the time convolution, since the kernel depends on the time history of quantities at the interface, which is only found during the time-stepping process. 
To address this, we develop a novel, causal version of the Fast Multipole Method that 
reduces the computational cost from $O(P^2)$ for direct evaluation of the time convolution to  $O(P \log_2^2 P)$ per surface grid point, 
where $P$ is the number of time steps. In the bulk phase, the resulting method is mesh-free and 
provides an accurate solution to the fully coupled moving interface problem with soluble surfactant. 
The approach extends naturally to a broader class of advection-diffusion problems in the 
high P\'eclet number regime.

\end{abstract}


\end{frontmatter}

\newpage
\section{Introduction} \label{Intro}

Surfactants are surface-active agents that accumulate at fluid interfaces, where they reduce surface tension and influence interfacial dynamics. 
This occurs via Marangoni forces, i.e., surface tension gradients,   which drive fluid motion along the interface. Surfactants play a crucial role in microfluidic applications and lab-on-a-chip devices, where they regulate droplet formation, suppress coalescence, and facilitate molecular exchange between droplets \cite{anna2016droplets}, \cite{baret2012surfactants}. Surfactants are also used in many processing applications as emulsifiers, detergents, and foaming or wetting agents. An overview of their properties and applications can be found in \cite{brenner2013interfacial}, \cite{manikantan2020surfactant}.

Insoluble surfactants are confined to the interface between immiscible fluids. 
In contrast, soluble surfactants exchange dynamically between their adsorbed form at the interface and their dissolved form in the neighboring bulk fluid, where they are transported by advection and diffusion.

In the zero Reynolds number or Stokes flow regime, interfacial evolution in the presence of an insoluble surfactant can be described by a Green's function formulation, which enables the use of boundary integral numerical methods \cite{pozrikidis1992boundary}. These are among the most accurate and efficient methods for solving Stokes flow problems in complex geometries. However, their extension to soluble surfactants is not straightforward because of the absence of a Green’s function formulation for the advection-diffusion equation governing surfactant transport in the bulk fluid.

A further challenge occurs  in  practice  because the bulk P\'eclet number $Pe$, which is the ratio of  characteristic advection to diffusion effects, is large, typically ranging from 
$10^5$ to $10^7$  \cite{chang1995adsorption, Moyle:2012, palaparthi2006theory}. At large P\'eclet numbers, surfactant transport in the bulk phase is dominated by advection, resulting in the formation of a thin transition layer adjacent to the interface where the surfactant concentration varies rapidly \cite{booty2010hybrid}. Resolving this transition layer is a significant challenge for standard numerical methods, but is essential for accurately capturing surfactant transport and exchange between the bulk phase and the interface.

Several numerical methods have been developed to simulate moving interfaces in the presence of soluble surfactant.
These include
a finite-difference/front-tracking algorithm for interfaces in 3D Navier-Stokes flow \cite{muradoglu2008front},   and an embedded boundary method for 2D Navier-Stokes flow  \cite{khatri2014embedded}.  Both approaches are demonstrated to be capable of handling moderate bulk P\'eclet numbers of  $Pe \simeq 10^2$.
Other  examples of numerical studies of drop dynamics with a soluble surfactant that are restricted to artificially small,  or at most moderate values of $Pe$,
are
\cite{ChenLai:2014, hu2018coupled, McLaughlin:2000, Milliken:1994, Teigen:2011, xu2018, xu2014, Zhang:2006,van2006diffuse,jin2006detachment,wang1999increased}.
 However, this P\'eclet number range is significantly lower than the  values found in typical applications.

 In this paper, we develop a fast and accurate boundary integral algorithm to solve the full-moving boundary problem for drops and bubbles with soluble surfactant at large $Pe$. Our approach builds on a ‘hybrid’ numerical method for the study of surfactant solubility effects in interfacial flow that was first introduced in \cite{booty2010hybrid} and extended in  \cite{xu2013analytical}, \cite{wang2014numerical}, \cite{wrobel2018simulation}, \cite{atwater2020studies}. 
 The method employs a singular perturbation analysis of the transition layer dynamics as  $Pe  \rightarrow \infty$ to derive an effective advection-diffusion equation that eliminates the difficulties associated with scale separation. The numerical solution of this equation is coupled to an accurate boundary integral simulation of the  dynamics at the drop interface, resulting in a multiscale method that captures the coupled evolution of fluid flow and bulk-interface surfactant transport. 
The algorithm is specifically designed for large P\'eclet numbers, corresponding to small bulk diffusion, and accurately resolves the narrow transition layer adjacent to the interface, where bulk surfactant concentration gradients normal to the interface are large.

 The efficiency and accuracy of the hybrid method was demonstrated in \cite{booty2010hybrid} by comparing it with a traditional numerical approach that uses finite differences on a curvilinear coordinate mesh without the transition layer reduction.  There have been improvements or refinements to the hybrid method, including an extension to axisymmetric geometry, a more general, mixed-kinetic boundary condition for the bulk-interface surfactant exchange, and a far-field condition for the bulk surfactant concentration that better resolves the structure of the transition layer \cite{wang2014numerical}. This method was adapted in \cite{wrobel2018simulation} to model a flow-focusing geometry and tip streaming phenomena observed in experiments \cite{anna2006microscale}. It has recently been extended to treat soluble surfactant in the drop interior \cite{atwater2020studies}, in which an internal recirculating flow causes bulk surfactant to advect in and out of an interior transition layer.

Previous implementations of the hybrid method have used  a mesh-based finite difference or Fourier-Chebyshev  approach
to solve for the  bulk surfactant concentration $C(\alpha,N,t)$ within a rectangular region  
$\alpha \times N \in  [0,2\pi] \times [0, N_{\text{max}}]$ that corresponds to the drop transition layer.
Here, \( \alpha \) denotes a normalized surface arc length parameter,  
and \( N \) is scaled distance normal to the interface.  
While this approach has the benefit of simplicity,  it has two  drawbacks:  
(i) The transition layer domain must be artificially truncated at some  \( N_{\text{max}} \), which introduces a potential source of error, and 
 (ii) resolving $C(\alpha,N,t)$  often requires  a large number of grid points  
 in the normal direction, which increases computational cost, although notably the asymptotic separation of scales implies that the number of grid points does not depend on \( Pe \).

In this paper, these limitations are addressed by exploiting the fact that, while the original advection-diffusion equation governing surfactant transport in the bulk fluid  does not have a Green’s function formulation, surprisingly, the transition layer equation admits one.   This formulation, derived in \cite{xu2013analytical},
gives the net flux of surfactant entering or leaving each interface point in terms of an Abel-type time-convolution integral with the form
\begin{equation} \label{eq:intro}
\left. \frac{\p C}{\p N} \right|_{N=0} = \int_0^{t} \frac{1}{\sqrt{t-\tau}} k(t,\tau) g(\tau) \,d\tau.
\end{equation}
The surfactant exchange flux (\ref{eq:intro})  appears as a source term in the equation for conservation of the surface concentration of surfactant. This provides
the foundation for a fully surface-based boundary integral method for the interfacial flow problem, including soluble surfactant.  This method was first presented in \cite{xu2013analytical},  and is mesh-free in the sense that no spatial mesh needs to be introduced in the direction normal to the interface to solve the transition layer equation for bulk surfactant concentration.  Instead, the spatial structure of the transition layer equation in the normal direction is contained in the convolution integral, while the tangential structure enters parametrically via a tracking Lagrangian fluid marker on the interface.

A further challenge in computing the bulk-interface surfactant exchange term (\ref{eq:intro})  
is the fast, i.e., efficient, evaluation of the time-convolution integral, which must be computed at every interface  
point and time step. A direct discretization of the integral with \( P \) time steps results in a  
computational cost of \( O(P^2) \) per surface grid point, making it expensive  
for large-scale simulations.    
The present study focuses on development and validation of a fast algorithm that reduces the cost to  
\( O(P \log_2^2 P) \) per surface grid point. The design of this algorithm must overcome  
two major difficulties:  (i) the complex structure of the kernel and (ii) the fact that the kernel is not known in advance—it depends on the complete time history of the quantities at the evolving interface, which is revealed only during the time-stepping process. 

To overcome these difficulties, we develop a new causal version of the Fast Multipole Method (FMM) that accelerates the computation. This approach was first introduced for the heat equation in \cite{tausch2007fast} and later extended to Volterra integral operators in \cite{tausch2012fast}, but significant modifications are required here to handle the nonstandard causality of the kernel.  
The resulting boundary integral algorithm for soluble surfactant is spectrally accurate in space and, with the incorporation  
of a standard FMM for spatial-convolution integrals, can attain  
an overall complexity over $P$ time steps of \( O(N_s P \log_2^2 P) \), where \( N_s \) is the number of interface  
grid points. However, in our implementation spatial convolutions are computed by  
a direct method, so that the current approach is validated by comparison with the earlier mesh-based method for soluble surfactant \cite{xu2013analytical}.  Several examples are presented.

Some simplifying assumptions have been made to facilitate the implementation of the method. We consider a single drop in a 2D geometry and assume that the soluble surfactant is present only in the exterior fluid. Away from the transition layer, the bulk surfactant concentration is treated as uniform. We impose a boundary condition for the bulk surfactant concentration at the interface that assumes the limit of diffusion-controlled kinetics rather than a more general, mixed-kinetic boundary condition. However, we note that modifying the method to handle 3D geometries, multiple drops, surfactant in the interior, nonuniform concentrations away from the transition layer, and mixed-kinetic boundary conditions is, in principle, straightforward.

The remainder of the paper is organized as follows. Section \ref{form} presents the full governing equations along with their reduced form in the infinite bulk Péclet number limit. Section \ref{sec:BI} introduces the boundary integral formulation, including the Green’s function representation of 
the bulk surfactant concentration within the transition layer. The numerical method, including the fast mesh-free algorithm for soluble surfactant, is developed in Section \ref{Surf Numerical Methods} and validated in Section \ref{OoA}.  Numerical examples are presented in Section \ref{sec:NumEx}. Concluding remarks are given in Section \ref{sec:Conclusion}.  Appendix A contains a validation of the fast mesh-free method using synthetic data.

\section{Governing equations} \label{form}

\begin{figure}[H]
    \centering
    \includegraphics[scale = 0.3]{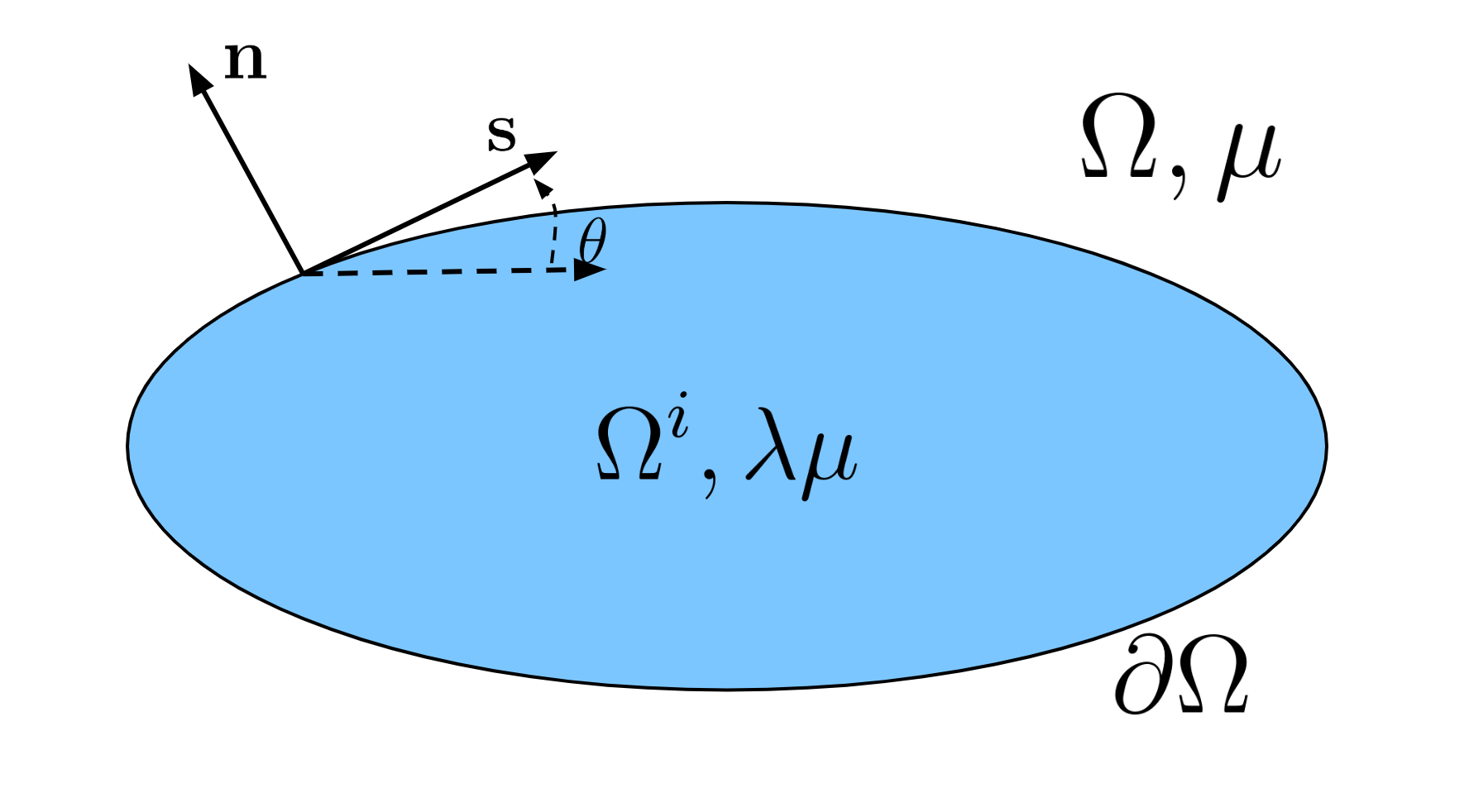}
    \caption{The 2D geometry: a drop of viscosity $\lambda \mu$ is surrounded by a fluid of viscosity $\mu$.  Shown are the unit outward normal vector $\mathbf{n}$, unit tangent vector   $\mathbf{s}$,  and the angle $\theta$ measured counterclockwise positive from the positive $x$-axis to $\mathbf{s}$.}
    \label{Formbubble}
\end{figure}

We consider an isolated fluid drop immersed in an infinite exterior fluid and deformed by an imposed flow. Let $\Omega$ denote the region exterior to the drop and $\Omega^i$ the region interior to the drop, where throughout a superscript $i$ is used to specify quantities in the drop interior. The exterior fluid has viscosity $\mu$ while the interior fluid has viscosity $\mu^i = \lambda\mu$.  The interface or boundary of the drop is denoted by  $\partial\Omega$.   The geometry is 2D, with additional notation illustrated in Figure \ref{Formbubble}.

The governing equations are presented in dimensionless form.
Lengths are nondimensionalized using the radius $a_0$ of the undeformed drop, and  surface tension is scaled by   $\sigma_0$, the value for a clean, surfactant-free interface. Velocities are nondimensionalized by the capillary velocity $U=\sigma_0/\mu$, time by the timescale $T_0=a_0/U$, and fluid pressure by the drop's capillary pressure $\sigma_0/a_0$.    The interfacial surfactant concentration  $\Gamma$ is nondimensionalized by the maximum monolayer packing concentration $\Gamma_\infty$, and the bulk surfactant concentration $C$ is scaled by a uniform far-field reference value $C_\infty$.  
The concentrations $\Gamma$ and $C$ represent distinct surfactant phases, and are assumed to be continuous in their respective domains for all $t>0$.

We present the main equations without derivation; additional  details can be found in \cite{xu2013analytical}.  In the zero-Reynolds number limit, the  fluid motion is governed by the  incompressible Stokes flow equations: 
\begin{equation}\label{Stokes}
\begin{aligned}
    \nabla^2\mathbf{u} =\nabla p,~~\nabla \cdot \mathbf{u} = 0, ~~\mathbf{x}\in\Omega, \\
    \lambda\nabla^2\mathbf{u}^i =\nabla p^i,~~\nabla \cdot \mathbf{u}^i = 0, ~~\mathbf{x}\in\Omega^i.
    \end{aligned}
\end{equation}
Here $\mathbf{u}$   is the fluid velocity in the exterior region,  $p$ denotes the pressure, and the superscript $i$ denotes the corresponding quantities in the interior fluid.  The fluid velocity is continuous across the interface, i.e., 
\begin{equation}\label{vinterface}
    \mathbf{u}(\mathbf{x},t) = \mathbf{u}^i(\mathbf{x},t), ~~\mathbf{x}\in\partial\Omega.
\end{equation}
The interface shape evolves according to the kinematic condition, which is    
\begin{equation}\label{kincond}
    \frac{d\mathbf{x}}{dt}\cdot \mathbf{n} = \mathbf{u}\cdot\mathbf{n}\equiv u_n, ~~\mathbf{x}\in\partial\Omega.
\end{equation}
In particular, if  $\mathbf{x}$ is the position of a material point on the interface, then it  satisfies
\begin{equation}\label{material}
    \frac{d\mathbf{x}}{dt} = \mathbf{u}(\mathbf{x},t),~~ \mathbf{x}\in \partial\Omega.
\end{equation}
In the inviscid limit $\lambda=0$, the interior pressure $p^i(t)$ is spatially constant, and the interior velocity $\mathbf{u}^i$ is not specified. The continuity condition (\ref{vinterface}) is then omitted, while the kinematic condition (\ref{kincond}) is retained.

The jump in fluid stress across the boundary is balanced by the interfacial surface tension. This gives the stress-balance boundary condition

\begin{equation}\label{stressbal}
    -(p-p^i)\mathbf{n} + 2(\mathbf{e}-\lambda\mathbf{e}^i)\cdot\mathbf{n} = \sigma\kappa\mathbf{n} - \nabla_s\sigma \text{,  } \mathbf{x}\in\partial\Omega,
\end{equation}
where $\kappa=-\partial \theta/\partial s$ is the curvature of the drop (taken positive where the shape is convex), with $\theta$ the angle shown in Figure \ref{Formbubble}, $\sigma$ is the surface tension, and $\mathbf{e}$ is the rate of strain tensor,  $e_{ij} = \frac{1}{2}(\partial_{x_i}u_j + \partial_{x_j}u_i) $. 
The stress-balance boundary condition couples the surfactant concentration to the fluid dynamics through the surface equation of state, $\sigma=\sigma(\Gamma)$.   
We adopt a linearized equation of state

\begin{equation}\label{eqstate}
    \sigma = 1-E\Gamma,
\end{equation}
where $E=R T\Gamma_\infty/\sigma_0$ is the elasticity parameter \cite{chang1995adsorption}.  This equation of state is usually employed for small surfactant concentrations, but for simplicity, we employ it here even for relatively large concentrations. A nonlinear equation of state, such as the Frumkin equation \cite{chang1995adsorption}, could be substituted with only minor modifications.

Soluble surfactant is present in the bulk fluid at a concentration $C$.
We assume soluble surfactant is present only in the exterior fluid, which is often the situation of interest in applications.  
While the algorithm developed in this study can be extended to account for surfactant in the interior fluid, doing so introduces additional complexities \cite{atwater2020studies}. The bulk surfactant concentration
$C$  satisfies the advection-diffusion equation:

\begin{equation}\label{bulkC}
    \frac{\partial C}{\partial t} + \mathbf{u}\cdot \nabla C = \frac{1}{Pe}\nabla^2C,~~ \mathbf{x}\in\Omega. 
\end{equation}
where $Pe=U a_0/D$ is the bulk P\'eclet number, which is typically large, of order $10^6$ or more,  in applications, and $D$ is the bulk surfactant diffusion coefficient.  The surface concentration of surfactant $\Gamma$ satisfies the conservation law  \cite{wong1996surfactant}
\begin{equation}\label{surfGam}
 \left.   \frac{\partial\Gamma}{\partial t} \right|_\xi - \left. \frac{\partial\mathbf{X}}{\partial t} \right|_\xi\cdot\nabla_s\Gamma +\nabla_s\cdot (\Gamma \mathbf{u}_s)+\Gamma\kappa u_n = \frac{1}{Pe_s}\nabla^2_s\Gamma + J\mathbf{n}\cdot\nabla C|_{\partial\Omega}, ~~\mathbf{x}\in\partial\Omega,
\end{equation}
where $\mathbf{u}_s=(\mathbf{u} \cdot \mathbf{s}) \mathbf{s}$ is the projection of the fluid velocity at the interface onto the surface tangent plane and $u_n=\mathbf{u}\cdot\mathbf{n}$ is the normal component of the fluid velocity at the interface, $\mathbf{x}=\mathbf{X}(\mathbf{\xi},t)$ is a parametric representation of the interface with curvature $\kappa$, $J=D C_\infty/U \Gamma_\infty$ is a measure of the diffusive flux of bulk surfactant relative to the advective flux of interfacial surfactant, and $Pe_s = Ua_0/D_s$ is the surface P\'eclet number with $D_s$ the surface diffusion coefficient of surfactant.

At the interface  we impose a Dirichlet boundary condition 
\begin{equation}\label{BCbulkC}
    C|_{\partial\Omega} = \frac{\Gamma}{K(1-\Gamma)},~~ \mathbf{x}\in\partial\Omega,
\end{equation}
where $K=\kappa_a C_\infty/\kappa_d$ is an equilibrium partition coefficient. Here, $\kappa_a$  and $\kappa_d$ denote the adsorption and desorption rates, respectively,  governing the transfer of surfactant between its adsorbed phase on the interface and its dissolved phase in the bulk.  Equation (\ref{BCbulkC}) is simplified from the more general flux relation 
\begin{equation*}
    J\mathbf{n}\cdot\nabla C|_{\partial\Omega} = Bi\Big(KC|_{\partial \Omega}(1-\Gamma)-\Gamma\Big),
\end{equation*}
by taking the infinite Biot number limit, $Bi \rightarrow \infty$, which is realistic in applications involving $0.1$ mm or larger sized drops. Here
$Bi=\kappa_d a_0/U$ is the ratio of the time scale of the flow to the time scale of the kinetic desorption process $\kappa_d^{-1}$. 
The right hand side $Bi(KC|_{\partial\Omega}(1-\Gamma)-\Gamma)$ is the kinetic relation corresponding to the Langmuir isotherm \cite{chang1995adsorption}, which expresses the rate of adsorption onto the interface minus the rate of desorption.

We prescribe a spatially uniform initial distribution of bulk surfactant, with a constant far-field value
\begin{equation}\label{ICbulkC}
    C(\mathbf{x},0)=1,~~ \mbox{for}~~\mathbf{x}\in\Omega; ~~C(\mathbf{x},t)\rightarrow1 \text{ as } |\mathbf{x}|\rightarrow\infty\text{ for } t>0.
\end{equation}
The initial distribution of interfacial surfactant is also chosen to be spatially uniform
\begin{equation}\label{ICbulkC2}
    \Gamma(\mathbf{x},0)= \Gamma_0,~~ \mbox{for}~~\mathbf{x}\in \partial \Omega.
\end{equation}
A linear far-field flow deforms the drop
\begin{equation}\label{flow}
    \mathbf{u}^\infty \sim 
    \left( 
    \begin{array}{c c}
    Q &  B+G/2 \\
    B-G/2  & -Q
    \end{array}
    \right)\cdot\mathbf{x}+O\left(\mathbf{|x|^{-2}}\right) \text{ as } |\mathbf{x}|\rightarrow\infty
\end{equation}
Here $Q, B,$ and $G$ are dimensionless parameters that equal their dimensional counterparts times the time scale $a_0/U$.  We focus on two canonical examples, (i) shear flow, where $\mathbf{u}^\infty\sim G(x_2,0)$ with $Q=0$ and $B=G/2$, and (ii) strain flow, where $\mathbf{u}^\infty\sim Q(x_1,-x_2)$ with $B=G=0$.

\subsection{The infinite bulk P\'eclet number limit} \label{sec:InfPeclet}
In the limit  $Pe\rightarrow\infty$, equation (\ref{bulkC}) for the bulk surfactant concentration is singularly perturbed. A narrow transition layer can form near the interface, where the normal gradient of  $C$  is large.  Outside this layer, the leading order concentration remains uniform provided it is initially uniform, which we assume here.   Using matched asymptotic expansions in  \cite{booty2010hybrid}, we derived the leading order evolution equation for $C$  in the transition layer as $Pe \rightarrow \infty$. This equation  provides an approximation of  the behavior of 
$C$  in the transition layer for large but  finite $Pe$ and is exact in the limit 
$Pe \rightarrow \infty$.   

The equation for $C$ in the transition layer is written in terms of a surface-fitted orthogonal coordinate system $(\xi,n)$ that is attached to and moves with the surface $\partial \Omega$ for $t>0$.  Here $\xi$ is aligned with the surface tangent and  $n$ measures distance from $\partial \Omega$ in the normal direction.  In the transition layer,  the position of a point $\mathbf{x}$ and fluid velocity $\mathbf{u}$ defined with respect to a fixed Eulerian coordinate system  are   decomposed as 
\begin{align}
\mathbf{x}&=\mathbf{X}(\xi,t) + n \mathbf{n} (\xi,t) \\
\mathbf{u}&=\mathbf{u}_t + u_p \mathbf{n}
\end{align}
where $\bx=\bXX(\xi,t)$  is the equation of $\partial \Omega$, $\bu_t$ is the projection of $\bu$ onto the surface tangent vector, and $u_p$ is  the  component of $\bu$  in the normal direction.  We also define  $\bUU_s= \partial \bXX/\partial t - \partial \bXX/\partial t  \cdot \bn$, which is the tangential velocity of an interface point at fixed $\xi$, and $\bv=\bv_t + v_p \bn$ with $\bv_t=\bu_t-\bUU_s$ and $v_p=u_p-u_n$, which is the fluid velocity relative to $\mathbf{X}(\xi,t)$.   In  the limit as the interface is approached,  $\bu_t$ tends to $\bu_s$,    the tangential fluid velocity at the interface, and  $v_p$ tends to zero.

 In the transition layer, $C$   depends on a local normal coordinate $N$, where $n =  \epsilon N$,  $\epsilon =  Pe^{-1/2}$ is the width of the transition layer, and $N = O(1)$   as $\epsilon \rightarrow 0$.  The leading order equation for $C$ in an expansion for small $\epsilon$ is \cite{booty2010hybrid}
\begin{equation}\label{bulkCN}
 \left.  \left.  \frac{\partial C}{\partial t} \right|_{\xi}  + \mathbf{v}_s\cdot \nabla C + \frac{\partial v_p}{\partial n} \right|_{\partial\Omega}N\frac{\partial C}{\partial N}= \frac{\partial^2 C}{\partial N^2},~~\bx \in \Omega^r
\end{equation}
where $\bv_s=\bu_s-\bUU_s$ is the tangential fluid velocity at the interface relative to an interface point with fixed $\xi$, and $\Omega^r$ is the transition layer subdomain of $\Omega$.  The error in approximating $C$ by the solution to  this equation is $O(\epsilon)$ as $\epsilon\rightarrow 0$.

Away from the transition layer, $C$ satisfies  $(\partial_t+\bu \cdot \nabla) C=0$ to the same order $O(\epsilon)$, so that $C$ is conserved on particle paths.  Therefore, (\ref{ICbulkC})  implies that $C=1$ in this outer region for $t>0$  and provides initial and matching conditions
\begin{equation}\label{BCbulkCN}
    C(\xi,N,0) = 1, ~~  C(\xi,N,t)\rightarrow1 \text{ as } N\rightarrow\infty\text{ for } t>0.
\end{equation}
The boundary condition for $C$ at the interface is given by (\ref{BCbulkC}),
\begin{equation} \label{eqn:C_BC}
    C|_{N=0} = \frac{\Gamma}{K(1-\Gamma)}.
\end{equation}
The incompressibility condition $\nabla \cdot \bu=0$ written in the intrinsic frame $(\xi,n)$ provides an expression for the coefficient  $\left. \partial_n v_p \right|_{\partial \Omega}$ in (\ref{bulkCN}) in terms of surface data alone
\be \label{vp_deriv_surf_form}
\left. \frac{\p v_p}{\p n} \right|_{\partial \Omega} = -\kappa u_n -  \nabla_s \cdot \bu_s.
\ee
Therefore, all the coefficients in (\ref{bulkCN}) can be evaluated from surface data, which is an important component of this method. 

The bulk-interface exchange term in equation (\ref{surfGam}) is similarly rescaled, so that the equation governing the surface concentration of surfactant becomes 
\begin{equation}\label{surfGamN}
\left.   \left.  \left. \frac{\partial\Gamma}{\partial t}\right|_\xi -\frac{\partial\mathbf{X}}{\partial t}\right|_\xi\cdot\nabla_s\Gamma +\nabla_s\cdot (\Gamma \mathbf{u}_s)+\Gamma\kappa u_n = \frac{1}{Pe_s}\nabla^2_s\Gamma + J_0\frac{\partial C}{\partial N}\right|_{\partial\Omega},~~ \mathbf{x}\in\partial\Omega. 
\end{equation}
Here $J_0$ is a rescaled transfer coefficient given by  $J=\epsilon J_0$ with $J_0=O(1)$.

We note that the expansion parameter $Pe$ does not appear in the initial boundary value problem (\ref{bulkCN})-(\ref{surfGamN}) for $C$ and $\Gamma$.  This is a consequence of retaining only leading order quantities in the limit $Pe \rightarrow \infty$.   The problematic scale separation that is present in the original formulation of the governing equations is now removed. 

\subsection{Arc-length--angle formulation of the governing equations} \label{sec:arclength-angle}

Following Hou, Lowengrub, and Shelley \cite{hou1994removing} and Kropinski \cite{kropinski2001efficient} we reformulate the governing equations using an arc-length--angle representation $s(\alpha,t),$ $\theta(\alpha,t)$ in which 
\begin{equation}\label{parametrization}
(x_{1\alpha}(\alpha,t),x_{2\alpha}(\alpha,t)) = s_\alpha(\alpha,t)(\cos\theta(\alpha,t),\sin\theta(\alpha,t))
\end{equation}
defines the associated Cartesian interface shape. Here the spatial parameterization of the interface is given by $\alpha \in [0, 2 \pi)$, $s(\alpha,t)$ is the arc length from a reference point at $\alpha=0$ to the point at $(x_1(\alpha,t),x_2(\alpha,t))$, $\theta(\alpha,t)$ is the angle shown in Figure \ref{Formbubble}, and the $\alpha$ subscript denotes $\partial_\alpha$. The arc length angle representation was originally developed to remove numerical stiffness, but we utilize it here as a convenient representation of the interface. 

The evolution equations for $s(\alpha,t)$ and $\theta(\alpha,t)$ are \cite{hou1994removing}
\begin{equation}\label{salpha}
    s_{\alpha t}  = (\phi_s)_\alpha -u_n\theta_\alpha
\end{equation}
and
\begin{equation}\label{theta}
    \theta_t = \frac{1}{s_\alpha}((u_n)_\alpha + \phi_s\theta_\alpha),
\end{equation}
where $u_n$ is the normal component of the fluid velocity at the interface and $\phi_s$ is the tangential velocity of a point on the interface with velocity 
\be \label{u_tang_norm_form}
{\bf{u}}=u_n {\bf{n}}+\phi_s{ \bf{s}}~~\mbox{for} ~\bf{x} \in \p \Omega. 
\ee
Equations (\ref{salpha}) and (\ref{theta}) are found by differentiating the kinematic condition (\ref{kincond})  with respect to $\alpha$, and equating normal and tangential components between the result and the equivalent representation (\ref{parametrization}).  While the normal velocity $u_n$ determines the interface shape,
the tangential velocity of interface markers $\phi_s$ can be chosen to specify the surface parametrization. 
A boundary integral representation for $u_n$, along with relations for  $\phi_s$ that enforce favorable parameterizations are given in Section \ref{sec:BI}. 
 
The Cartesian coordinates $(x_1,x_2)$ of the interface are recovered from $(s_\alpha, \theta)$ by
\begin{equation}\label{cartesian}
    \begin{aligned}
        x_1(\alpha,t) = x_1(0,t) + \int^\alpha_0 s_\alpha(\alpha ',t)\cos(\theta(\alpha',t))d\alpha' \\
        x_2(\alpha,t) = x_2(0,t) + \int^\alpha_0 s_\alpha(\alpha ',t)\sin(\theta(\alpha',t))d\alpha',
    \end{aligned}
\end{equation}
which are obtained by integrating \eqref{parametrization} with respect to $\alpha$. The position of the reference point $\alpha=0$ at time $t$, ($x_1(0,t)$,$x_2(0,t)$), is time-evolved from (\ref{u_tang_norm_form}). 

\section{Boundary integral formulation} \label{sec:BI}

We use the Sherman-Lauricella representation as our boundary integral formulation, which has been widely used to describe $2$D boundary value problems involving the biharmonic equation in contexts such as  linear elasticity and Stokes flow, see e.g., \cite{greengard1996integral}, \cite{kropinski2001efficient}, \cite{mikhlin2014integral}, \cite{xu2013analytical}. The presentation here follows \cite{xu2013analytical}.  In this formulation,  the primitive variables are expressed in terms of a single complex density $\omega$ that is defined on the drop interface and satisfies a Fredholm second kind integral equation.  The equation for the density is 
\begin{equation}\label{SLdensity}
\begin{aligned}
    \omega(\eta,t) +\frac{\beta}{2\pi i}\int_{\partial\Omega}\omega(\zeta,t)d\ln{\frac{\zeta-\eta}{\overline{\zeta}-\overline{\eta}}} +& \frac{\beta}{2\pi i}\int_{\partial\Omega}\overline{\omega(\zeta,t)}d\frac{\zeta-\eta}{\overline{\zeta}-\overline{\eta}}\\&+\beta(B-i Q)\overline{\eta}+2\beta H(t)
    = -\frac{\rho}{2}\sigma(\Gamma)\frac{\partial\eta}{\partial s},
    \end{aligned}
\end{equation}
where
\begin{equation*}
    \beta = \frac{1-\lambda}{1+\lambda},~~ \rho = \frac{1}{1+\lambda},
\end{equation*}
and $\eta=x_1+ix_2$ is a point on $\partial \Omega$ in complex format.   The integral equation (\ref{SLdensity}) is rank deficient in the limit $\lambda=0$, and the choice 
\[
H(t)=\frac{1}{2} \int_{\p \Omega}  \omega(\zeta,t) ~ ds
\]
removes the deficiency where $H(t) \equiv 0$  as a consequence of the constant area of the interior drop region \cite{kropinski2001efficient}. We note that the apparent singularity at $\zeta=\eta$ in
the two integrals on the left-hand side is removable. 
The fluid velocity $\mathbf{u}$ on the interface is written  in complex form $u|_{\partial\Omega} = u_1+iu_2$, and  is given in terms of the density $\omega$ by
\begin{equation}\label{SLvelo}
\begin{aligned}
    u_1+i u_2|_{\partial\Omega} &= -\frac{1}{2\pi}\dashint_{\partial\Omega} \omega(\zeta,t)(\frac{d\zeta}{\zeta-\eta}+\frac{d\overline{\zeta}}{\overline{\zeta}-\overline{\eta}})\\ &+ \frac{1}{2\pi}\int_{\partial\Omega}\overline{\omega(\zeta,t)}d(\frac{\zeta-\eta}{\overline{\zeta}-\overline{\eta}}) + (Q+i B)\overline{\eta}-\frac{i G}{2}\eta.
    \end{aligned}
\end{equation}
The second integral in equation \eqref{SLvelo} has a removable singularity when  $\zeta=\eta$, while the first integral is a Cauchy principal value integral.  

For the mesh-free numerical method developed here we utilize a material, i.e.,  Lagrangian, parameterization of the interface $\p \Omega$, for reasons which will be apparent in Section \ref{GF}.    The velocity of a point $\eta$  on the interface is then equal to the local fluid velocity, so that in  (\ref{u_tang_norm_form}) the tangential interface velocity is set as  $\phi_s =u_s$ where $u_s={\bf u} \cdot {\bf s}$ at ${\bf x} \in \p \Omega$. It follows that 
\be
\frac{d \eta}{d t} = u_n i e^{i \theta} + u_s e^{i \theta} \nonumber
\ee
which is obtained from (\ref{u_tang_norm_form}) using
the complex form of  the unit normal  and tangent vectors, i.e.,
\[
u_n=Re \left[  \left. (u_1 + i u_2) \right|_{\partial \Omega} \bar{n}  \right]~~~u_s=-Im \left[  \left. (u_1 + i u_2) \right|_{\partial \Omega} \bar{n}  \right].
\]
The mesh-free method will be compared to the previously developed mesh-based method of \cite{booty2010hybrid} and \cite{xu2013analytical}. The mesh-based approach employs an equal-arc-length  parameterization of the interface, in which 
\[
s_\alpha(\alpha,t) = l(t),
\]
 so that  $\p s / \p \alpha$ is  independent of $\alpha$ but depends on time.  The tangential interface velocity $\phi_s$  required to maintain the  equal-arc--length parameterization is 
 given by 
\be \label{phi_s_equal_arclength}
\phi_s (\alpha,t) = -\frac{\alpha}{2 \pi}  \int_0^{2 \pi} u_n \theta_{\alpha'}  \ d \alpha' + \int_0^\alpha  u_n \theta_{\alpha'} \ d \alpha'.
\ee
This formulation ensures uniform point distribution along the interface and follows the original approach described in \cite{hou1994removing}.

\subsection{Green's function solution for the transition layer}\label{GF}

In  \cite{xu2013analytical},  Xu et al.\  express the solution of the transition layer equation (\ref{bulkCN})  with (\ref{BCbulkCN}) to (\ref{vp_deriv_surf_form}) using a Green's function, which gives a boundary integral formulation for the complete Stokes flow problem with soluble surfactant as $Pe \rightarrow \infty$.  The surface parameterization is such that $\alpha \in [0,2 \pi)$ is a Lagrangian coordinate on the interface, and the bulk surfactant concentration in the transition layer $C(\alpha,N,t)$ is taken to be  $2 \pi$-periodic in $\alpha$.  Equations  (\ref{bulkCN}) to (\ref{vp_deriv_surf_form}) are first written as
\begin{equation}\label{bulkCMesh}
    \frac{\partial C}{\partial t} + v_s(\alpha,t)\frac{\partial C}{\partial s} + \psi(\alpha,t)N\frac{\partial C}{\partial N}= \frac{\partial^2 C}{\partial N^2}
\end{equation}
where 
\begin{equation}  \label{psiMesh}
    v_s(\alpha,t) = u_s(\alpha,t)-\phi_s(\alpha,t), ~~~ \mbox{and}~~~ \psi(\alpha,t) = -(\kappa u_n+\frac{\partial u_s}{\partial s}).
\end{equation}
The initial and matching  conditions (\ref{BCbulkCN})  are 
\be \label{BCMesh0}
C(\alpha,N,t=0)=1 ~~~\mbox{and}~~~C(\alpha,N,t)\rightarrow1 \text{ as } N\rightarrow\infty,
\ee
and the boundary condition (\ref{eqn:C_BC})   is
\begin{equation}\label{BCMesh}
     C(\alpha,N=0,t) = \frac{\Gamma(\alpha,t)}{K(1-\Gamma(\alpha,t))}.
\end{equation}

With $\alpha=s_0 \in [0,2 \pi)$ as the arc length of a Lagrangian fluid marker at $t=0$, its arc length $s$ at time $t>0$ is given by
introducing characteristic paths
\begin{equation}\label{charpath}
    s = f(s_0,t), \text{ such that } \frac{\partial s}{\partial t}= v_s(s,t), \text{ with } s=s_0 \text{ at } t=0.
\end{equation}
In this coordinate frame we define 
\begin{equation}\label{psi0}
    \psi_0(t) = \psi(s=f(s_0,t),t), ~~\Gamma_0(t) = \Gamma(s=f(s_0,t),t),
\end{equation}
so that the transition layer equation (\ref{bulkCMesh}) becomes 
\begin{equation}\label{bulkCGF}
    \frac{\partial C}{\partial t} + \psi_0(t) N\frac{\partial C}{\partial N} = \frac{\partial^2 C}{\partial N^2},
\end{equation}
where we shift $C \mapsto 1+C$.
The initial and boundary conditions  are now
\be \label{Ceqn_init_BC}
C(N,t=0) = 0,~~ C(N=0,t) = h_0(t),~~\mbox{and}~~C(N\rightarrow\infty,t) = 0, 
\ee
where
\begin{equation}\label{h0}
    h_0(t) = \frac{\Gamma_0(t)}{K(1-\Gamma_0(t))}-1.
\end{equation}
Note that we have suppressed the spatial dependence of $h_0$, $\Gamma_0$ and $C$ on $s_0$.  This initial boundary value problem is solved in Xu et al.\  \cite{xu2013analytical} by Duhamel's principle with a similarity variable $N/\gamma$, which gives  the representation
\begin{equation}\label{Bulk C Solution}
    C(s_0,N,t)=1+\frac{2N}{\sqrt{\pi}}\int_{0}^{t}e^{-(\frac{N}{\gamma})^2}\frac{\partial_u\gamma}{\gamma^2}h_0(t-u)du,
\end{equation}
in which $\gamma=\gamma(u,t-u)$, where
\begin{equation}\label{gamdcdn}
    \gamma(u,t-u) = 2e^{\frac{1}{2}a(u,t-u)}\Big(\int^{u}_0 e^{-a(\Tilde{t},t-u)}d\Tilde{t}\Big)^{1/2}~~\mbox{with}~~ a(t,\tau) = 2\int^t_0\psi_0(t'+\tau)dt'.
\end{equation} 
We observe that as $u \rightarrow 0$, the following expansions hold:
 \begin{equation}\label{gam and a u->0}
    \begin{aligned}
        \gamma(u,t-u) &= 2\sqrt{u}\Big(1+\frac{\psi_0(t)}{2}u+O(u^2)\Big),\\
        \partial_u\gamma(u,t-u)&=\frac{1}{\sqrt{u}}\Big(1+\frac{3\psi_0(t)}{2}u+O(u^2)\Big).
    \end{aligned}
\end{equation}
These expressions will be used in the subsequent analysis.

Equation \eqref{Bulk C Solution} is used to calculate the normal derivative  $\partial_N C|_{N=0}$ at the interface in the surfactant exchange term of (\ref{surfGamN}), but care must be taken in view of the singularity of $\gamma(u,t-u)$ as $u\rightarrow0$. The result is  \cite{xu2013analytical}
\begin{equation}\label{dcdn}
  \left.  \frac{\partial C}{\partial N}\right|_{N=0} = -\frac{2h_0(0)}{\sqrt{\pi}\gamma(t,0)} - \frac{2}{\sqrt{\pi}}\int^t_0 \frac{\partial_\tau h_0(\tau)}{\gamma(t-\tau,\tau)}d\tau.
\end{equation}
Note that $h_0(0)=0$ if the bulk and interface surfactant concentrations are initially in equilibrium, in which case the first term in  \eqref{dcdn} is zero.  The integrand in (\ref{dcdn})  has an integrable singularity as  $\tau\rightarrow t^-$  since,  from (\ref{gam and a u->0}),  $\gamma(t-\tau,\tau)\sim2(t-\tau)^\frac{1}{2}$ in this limit.  There is also a $\tau^{-\frac{1}{2}}$ singularity in $\partial_\tau h_0(\tau)$  as $\tau\rightarrow0^+$  if the initial bulk and interface concentration of surfactant are not in equilibrium. However,  when they are in equilibrium $h_0(\tau)\sim\tau$ as $\tau\rightarrow0$. 

Equation (\ref{dcdn})  allows the evaluation of the bulk-interface surfactant exchange term in (\ref{surfGamN}) without the explicit introduction of a mesh in the transition layer. 
The numerical method based on equation (\ref{dcdn}) is therefore referred to as `mesh-free.'

\section{Numerical Method}\label{Surf Numerical Methods}

The fast mesh-free method to compute the bulk-interface surfactant exchange term (\ref{dcdn})  and the bulk surfactant concentration  $C$ of (\ref{Bulk C Solution})  is described in Section \ref{FastHM}.
Apart from this
the numerical method mostly follows \cite{xu2013analytical}.  

The surface parameter $\alpha=s_0$ is discretized with a uniform step size as $\alpha_j  \in [0, 2 \pi)$ for $j=1, \ldots, N_s$.   Additionally, we assume that the interface shape, defined by $s_\alpha$ and  $\theta$  as well as the surfactant concentrations $\Gamma$ and $C$, are known at time $t=t_n$.  The update to time $t_{n+1}$ is obtained as follows. 

The interface velocity is computed from  (\ref{SLdensity}) and  (\ref{SLvelo}), with the smooth quadratures in (\ref{SLdensity}) evaluated by the trapezoidal rule,  
and the Cauchy principal value integral in (\ref{SLvelo}) computed by the Van de Vooren correction \cite{van1980numerical}, ensuring spectral accuracy in space. 
Time updates to $s_\alpha$ and $\theta$ are computed from the evolution equations (\ref{salpha}) and (\ref{theta}) using a second-order Adams-Bashforth method.  In the mesh-free method, a Lagrangian parameterization is employed so that  $\phi_s=u_s$. The method is validated by comparison with a previously developed
mesh-based method \cite{booty2010hybrid,xu2013analytical}, which employs an equal-arc--length parameterization, for  which $\phi_s$  is defined by (\ref{phi_s_equal_arclength}). 
The time update in equation (\ref{surfGamN}) for the evolution of  $\Gamma$ depends on the bulk-interface surfactant exchange term $\partial_N C|_{N=0}$, which introduces several subtleties in its implementation. The Cartesian coordinates $(x_1,x_2)$ of the interface are recovered from $(s_\alpha, \theta)$ following equation  (\ref{cartesian}). All spatial derivatives and integrals are evaluated explicitly via fast Fourier transforms (FFTs).

\subsection{Fast mesh-free method}\label{FastHM} 

A bottleneck in the computation of the bulk-interface surfactant exchange term  (\ref{dcdn}) is the evaluation of the convolution, which must be performed at each time step and at each  interface grid point $\alpha_j$ for $j=1, \ldots N_s$. If the integral is discretized using $P$ time steps $t_k$, $k=1, \ldots, P$, then naive computation of the convolution requires $O(P^2)$ operations per surface grid point.   Here, we present a fast algorithm that evaluates the convolution in $O(P \log_2^2 P)$ operations per surface grid point.  As already noted, two challenges in developing this algorithm are the complex form of the kernel, as defined in equations (\ref{gamdcdn}) to (\ref{dcdn}), and  the fact that the kernel is not known in advance, since it depends on the time history of quantities at the evolving interface and the interface curvature through the function $\psi_0(t)$, which is only found during the time-stepping process. 

\underline{\em{Reformulation of the kernel}}.  The kernel in (\ref{dcdn}) is represented in terms of integrals via the functions $a(t,\tau)$  and $\gamma(u,t-u)$ of  (\ref{gamdcdn}).  An alternative representation that simplifies the numerical evaluation of the integral operator in (\ref{dcdn}) is given by evaluation of a pair of ODEs as follows.
Define 
\begin{equation}\label{psi1}
    \psi_1(t) = \int_{0}^{t}\psi_0(\Tilde{t})d\Tilde{t},
\end{equation}
where $\psi_0$ is given by  \eqref{psiMesh} and \eqref{psi0}. Then equation \eqref{gamdcdn} becomes
\begin{equation}\label{gamsquare1}
    \gamma^2(t-\tau,\tau) = 4\exp[2\psi_1(t)]\int_{\tau}^{t}\exp[-2\psi_1(u)]du.
\end{equation}
Further, define
\begin{equation}\label{psi2}
    \psi_2(t) = \int_{0}^{t}\exp[-2\psi_1(u)]du,
\end{equation}
so that 
\begin{equation}\label{gamsquare2}
    \gamma^2(t-\tau,\tau) = 4 \exp\left[2\psi_1(t)\right] 
   \Big(\psi_2(t)-\psi_2(\tau) \Big).
\end{equation}
Then $\psi_1$ and $\psi_2$ satisfy the system of ordinary differential equations 
\begin{equation}\label{psiODE}
\begin{aligned}
    \psi_1'(t) &= \psi_0(t),\\
\psi_2'(t) &= \exp[-2 \psi_1(t)],
\end{aligned}
\end{equation}
with initial conditions $\psi_1(0)=\psi_2(0)=0$.  Thus the kernel of the integral operator in (\ref{dcdn}) is given by
\be \label{kernel1}
\frac{2}{\sqrt{\pi}} \frac{1}{\gamma(t-\tau,\tau)}= 
\frac{1}{\sqrt{\pi}} \frac{\exp[-\psi_1(t)]}{\sqrt{\psi_2(t)-\psi_2(\tau)}}.
\ee

We now write the integral operator as a generalized Abel integral.
To that end, introduce 
\begin{equation}\label{k(t,tau)}
    k(t,\tau) = \frac{2\sqrt{t-\tau}}{\sqrt{\pi}\gamma(t-\tau,\tau)} = \frac{1}{\sqrt{\pi}}\exp\left[-\psi_1(t)\right] \left(
\frac{t-\tau}{\psi_2(t)-\psi_2(\tau)} \right)^\frac{1}{2}.
\end{equation}
The ODE system (\ref{psiODE}) implies that $\psi_2(t)$ is monotonically increasing so that $k(t,\tau)$ is real-valued and is a smooth function in the closed triangle
\be \label{triangle_def}
\Delta_T = \left\{(t,\tau): 0 \leq \tau \leq t \leq T \right\}
\ee
with limiting values on the edge $\tau=t$ given by 
\be  \label{limiting_value}
k(t,t) = \frac{1}{\sqrt{\pi}} \frac{\exp[-\psi_1(t)]}{\sqrt{\psi_2'(t)}} = \frac{1}{\sqrt{\pi}}.
\ee
It follows from (\ref{kernel1}) that the convolution in (\ref{dcdn}) can be written as an Abel-type integral operator
\begin{equation}\label{kernel}
    \mathcal{K} g(t) = \int_0^t \frac{1}{\sqrt{t-\tau}} k(t,\tau)  g(\tau) \, d\tau,
\end{equation}
where $g(\cdot)$ is the time derivative of $h_0(\cdot)$.
In the case when the bulk and interfacial concentrations of surfactant are in equilibrium,  the first term on the right-hand side of \eqref{dcdn} is zero
and this equation simplifies to
\begin{equation}\label{kernel equilibrium}
 \left.   \frac{\partial C}{\partial N} \right|_{N=0}= -\mathcal{K} g(t).
\end{equation}

An unusual aspect of this particular Abel integral operator is that   $k(t,\tau)$
is not known a priori in $\Delta_T$ but is defined in terms of the solution on 
 $[0,t]$.  This feature has implications for the design of our fast numerical method.

 \underline{\em{Discretization of the integral operator}}. 
As noted below equation (\ref{dcdn}), the kernel in  \eqref{kernel} has a singularity of order  $-\half$ at $t=\tau$. There is also a singularity in $g(\tau)=\p_\tau h_0(\tau)$ at $\tau=0$, which is of the form 
\begin{equation}\label{h0ste}
 g(\tau)=   \partial_\tau h_0(\tau) = \frac{A_0}{\sqrt{\tau}} + A_1 + A_2\sqrt{\tau} + O(\tau).
\end{equation}
Expressions for $A_0$ and $A_1$ were determined by Taylor expansion in \cite{xu2010computational}  and are given by 
\begin{equation}\label{A0A1}
    \begin{aligned}
        A_0 = -\frac{J_0 h_0(0)}{\sqrt{\pi}K(1-\Gamma(0))^2}\\
        A_1 = \frac{1}{K(1-\Gamma(0))^2}\Tilde{R}(0)
    \end{aligned}
\end{equation}
where $J_0$ is defined after equation (\ref{surfGamN}), 
\begin{equation}\label{R(t)}
    \Tilde{R}(t) = \mathbf{u}_s\cdot\nabla_s\Gamma-\nabla_s\cdot (\Gamma \mathbf{u}_s)-\Gamma\kappa u_n +\frac{1}{Pe_s}\nabla^2_s\Gamma,
\end{equation}
and  $\left. \p {\mathbf X}/\p t \right|_\xi=\mathbf{u}_s$  for the characteristic paths (\ref{charpath}).
The expression  for $A_2$, derived by  Taylor expansion  in  \cite{samantha}, is
\begin{equation}\label{A2 surf}
    A_2 = \frac{1}{K(1-\Gamma(0))^2}\Big(\Tilde{R}_1 -\frac{2J_0\Tilde{R}(0)}{K\sqrt{\pi}(1-\Gamma(0))^2} \Big),
\end{equation}
where $\Tilde{R}_1$ is found from the small time expansion of $\Tilde{R}(t) = \Tilde{R}(0) + \Tilde{R}_1t^\half+ \Tilde{R}_2t + O(t^\frac{3}{2})$ and can be computed as
\begin{equation}\label{R(1) tilde}
    \Tilde{R}_1 = \frac{\Tilde{R}(\Delta t)-\Tilde{R}(0)}{\sqrt{\Delta t}} ~~\mbox{for}~~\Delta t \ll 1~~
\end{equation}
with error $O(\Delta t^{1/2})$.

A quadrature is needed that accurately evaluates both of the aforementioned singularities. To this end, we employ a modified trapezoidal rule with endpoint corrections based on the generalized Euler-Maclaurin formula. Quadrature rules of this type were first introduced by Navot   \cite{navot1961extension} and extended to Abel-type integrals by Tausch \cite{tausch2010generalized}. 
The following quadrature rule can be derived based on  \cite{tausch2010generalized} for sufficiently smooth $\varphi$:
\begin{multline}\label{discretize0}
%\mathcal{K} g(t)=
\mathcal{K} g(t_n)= \int_0^{t_n} (t_n-\tau)^{-\half} \varphi(t_n,\tau) \,d\tau  \\
= 
h \sum_{j=1}^{n-1} (t_n-t_j)^{-\half} \varphi(t_n,t_j) 
 +  h^\half \Big( r_0 \varphi(t_n,t_n) + r_1 \varphi(t_n,t_{n-1}) \Big) + O \left( h^{\frac{5}{2}} \right) 
%& + s_n^{(0)} \varphi_0(t_n) + s_n^{(1)} \varphi_1(t_n)  +  s_n^{(2)} \varphi_2(t_n)
\end{multline} 
where  $h$ is the time-step size, given by $t_n=nh$, and 
\be
r_0=\zeta \left( -\frac{1}{2} \right) - \zeta \left( \frac{1}{2} \right) ~~\mbox{and}~~r_1=-\zeta \left( -\frac{1}{2} \right). \nonumber
\ee
Here $\zeta(\cdot ) $ is the Riemann zeta function.  Higher-order discretizations can be derived following the prescription in \cite{tausch2010generalized}. We note that 
(\ref{discretize0}) is valid only when $\varphi$ and its first derivatve vanish at $\tau=0$. This does not hold for $\varphi(t, \tau ) = k(t, \tau) \p_\tau h_0 (\tau)$, but this can be resolved by the method of singularity subtraction. Let
\be \label{sing_subtract}
\hat{\varphi}(t, \tau) = \varphi(t, \tau) -\frac{\varphi_0(t)}{\sqrt{\tau}} -\varphi_1(t)-\sqrt{\tau} \varphi_2(t)-\tau \varphi_3(t) ,
\ee
with  $\hat{\varphi}(t, \tau)=O(\tau^{3/2})$ as $\tau \rightarrow 0$.  Then, applying (\ref{discretize0}) with $\varphi$ replaced by $\hat{\varphi}(t, \tau)$ and the singularity subtraction terms integrated analytically, leads to 
\begin{multline}  \label{discretize}
\mathcal{K} g(t_n)=\int_0^{t_n} (t_n-\tau)^{-\half} \varphi(t_n,\tau) \,d\tau  \\
= 
h \sum_{j=1}^{n-1} (t_n-t_j)^{-\half} \varphi(t_n,t_j) 
 +  h^\half \Big( r_0 \varphi(t_n,t_n) + r_1 \varphi(t_n,t_{n-1}) \Big) \\
 + s_n^{(0)} \varphi_0(t_n) + s_n^{(1)} \varphi_1(t_n)  +  s_n^{(2)} \varphi_2(t_n)  + s_n^{(3)} \varphi_3(t_n)+ O(h^{\frac{5}{2}}),
\end{multline}
where
\begin{equation}\label{sn}
    \begin{aligned}
  s_n^{(0)} &= B \left( \frac{1}{2}, \frac{1}{2} \right)
  - h \sum_{j=1}^{n-1} (t_n-t_j)^{-\half} t_j^{-\half}
  - h^{\half} \left( \frac{r_0}{\sqrt{t_n}} + \frac{r_1}{\sqrt{t_{n-1}}}\right), \\
  s_n^{(1)} &= \sqrt{t_n}   B \left( \frac{1}{2}, 1 \right)
                - h \sum_{j=1}^{n-1} (t_n-t_j)^{-\half} 
  - h^{\half}\left( r_0 + r_1\right),\\
  s_n^{(2)} &=  t_n  B \left( \frac{1}{2}, \frac{3}{2} \right)
                - h \sum_{j=1}^{n-1} (t_n-t_j)^{-\half}t_j^{\half}
  - h^{\half} \left( r_0\sqrt{t_n} + r_1\sqrt{t_{n-1}} \right),\\
  s_n^{(3)} &= t_n^{\threehalf} B \left(\half,2\right) - h
  \sum_{j=1}^{n-1}(t_n-t_j)^{-\half} t_j - h^{\half} \big(r_0 t_n + r_1
  t_{n-1} \big),
  \end{aligned} 
\end{equation}
and $B(\mu,\nu)$ is the Euler beta function
\be
B(\mu,\nu) = \int_0^1 (1-x)^{\mu-1} x^{\nu-1} \ dx =\frac{\Gamma(\mu) \Gamma(\mu)}{\Gamma(\nu+\mu)},
\ee
for which $B \left( \frac{1}{2}, \frac{1}{2} \right)=\pi$, $ B \left( \frac{1}{2}, 1 \right)=2$, $B \left( \frac{1}{2}, \frac{3}{2} \right)= \frac{\pi}{2}$, and $B\left(\half,2\right) = \frac{4}{3}$.
The quantities $s_n^{(k)}$ in (\ref{sn})  do not depend on $\varphi$ and are  computed {\em a priori}.

For the evaluation of (\ref{kernel}) it is challenging to
  determine $\varphi_3(t)$ in (\ref{sing_subtract}) and therefore it will be
  omitted in (\ref{discretize}).  The generalized Euler-Maclaurin expansion 
  \cite{navot1961extension} implies that
  $s_n^{(3)}= h^{\threehalf} [\frac{1}{12} n^{-\half} +
  O(n^{-\threehalf}) ]$ and therefore the error in (\ref{discretize}) without 
  $s_n^{(3)} \varphi_3(t)$  is $O(h^{\threehalf})$.
  
We  replace  the function $\varphi$  of (\ref{discretize})  with
\begin{equation}\label{varphi}
    k(t, \tau) \p_\tau h_0 (\tau)=\frac{ \exp[-\psi_1(t)]}{\sqrt{\pi}} 
  \left(\frac{t-\tau}{\psi_2(t)-\psi_2(\tau)} \right)^\half 
 \partial_\tau h_0(\tau).
\end{equation}
To compute the functions $\varphi_k(t)$ in (\ref{discretize}),  Taylor expand  (\ref{varphi}) about $\tau=0$  to  find
\begin{equation}\label{varphi2}
    \left(\frac{t-\tau}{\psi_2(t)-\psi_2(\tau)} \right)^\half
= k_0(t) + \tau k_1(t) + O(\tau^2),
\end{equation}
where
\begin{equation}\label{k0k1}
    k_0(t) = \left( \frac{t}{\psi_2(t)}\right)^\half\quad\mbox{and}\quad
  k_1(t) = \frac{t-\psi_2(t)}{2 t^\half \psi_2(t)^\threehalf},
\end{equation}
where we note that $\psi_2(\tau) \sim \tau$ as $\tau \rightarrow 0$.
Multiplication of  this expansion by \eqref{h0ste} gives 
\begin{equation}\label{varphi coeff}
    \begin{aligned}
  \varphi_0(t) &= k_0(t) A_0\\
  \varphi_1(t) &= A_1\\
  \varphi_2(t) &= A_0 k_1(t) + A_2 k_0(t).
\end{aligned}
\end{equation}
Note that (\ref{discretize}) is not defined for the first time step $t_1=h$. In this case, the integral can be evaluated using the expansion (\ref{sing_subtract}), i.e., 
\begin{multline} 
   \int_0^{h} (h-\tau)^{-\half} \varphi(h,\tau) \,d\tau  \\
= \int_0^h (h-\tau)^{-\half} \left(  \frac{\varphi_0(t)}{\sqrt{\tau}} +\varphi_1(t)+\sqrt{\tau} \varphi_2(t) \right) \, d\tau \\
+ \int_0^h (h-\tau)^{-\half} \hat{\varphi}(h,\tau) \,d\tau  \label{first_time_step} \\
 = B \left( \frac{1}{2}, \frac{1}{2}  \right) \varphi_0(h)  
 + h^{\frac{1}{2}} B 
 \left( \frac{1}{2}, 1 \right) \varphi_1(h)
  +  h B \left( \frac{1}{2}, \frac{3}{2} \right) \varphi_2(h) + O(h^{\frac{3}{2}}).
\end{multline}

\begin{remark}
As noted, omission of $s_n^{(3)} \varphi_3(t)$  in (\ref{discretize}) and (\ref{first_time_step})  increases  the quadrature error 
to $O(h^\frac{3}{2})$.
This could be improved to $O(h^\frac{5}{2})$ by retaining  $s_n^{(3)} \varphi_3(t)$,  or to higher-order accuracy by further singularity subtraction in (\ref{sing_subtract}).  
\end{remark}

\begin{figure}[h!]

\begin{center}
\setlength{\unitlength}{0.75cm}
\begin{picture}(9,9)
  \put(0,0){\vector(1,0){9}}
  \put(0,0){\vector(0,1){9}}
  \put(8,0){\line(0,1){8}}
  \put(0,0){\line(1,1){8}}
  \put(2,2){\line(1,0){6}}
  \put(2,2){\line(0,-1){2}}
  \put(4,4){\line(1,0){4}}
  \put(4,4){\line(0,-1){4}}
  \put(6,6){\line(1,0){2}}
  \put(6,6){\line(0,-1){6}}
  \put(1,1){\line(1,0){3}}
  \put(1,1){\line(0,-1){1}}
  \put(3,3){\line(1,0){3}}
  \put(3,3){\line(0,-1){3}}
  \put(5,5){\line(1,0){3}}
  \put(5,5){\line(0,-1){3}}
  \put(7,7){\line(1,0){1}}
  \put(7,7){\line(0,-1){3}}
  \put(8.6,0.3){\makebox(0,0){$t$}}
  \put(0.3,8.6){\makebox(0,0){$\tau$}}
  \inblue{\linethickness{1.2pt}\put(4.5,4.5){\line(0,-1){4.5}}}
\end{picture}
    \caption{Partition of triangle $\Delta_T$ for the fast evaluation of the convolution, showing two levels.  The kernel is known only in the subset $\Delta_t$ in which $\tau \leq t$, shown bounded on the right by the thick blue line.}
    \label{triangle}
\end{center}
\end{figure}
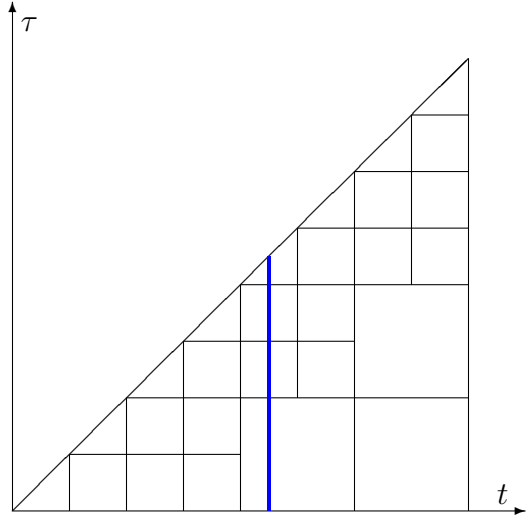

 \underline{\em{Fast computation of the convolution}}.
 The dominant computational cost of evaluating the discretization of $\K g$ at the
$n$-th time step is the evaluation of the trapezoidal sum in
\eqref{discretize}. For this type of operator, a causal version of the
Fast Multipole Method can be used to accelerate the computation. This approach was first introduced in the context
of the heat equation in ~\cite{tausch2007fast} and later extended to  Volterra integral
operators in~\cite{tausch2012fast}. The method presented there assumes that the
kernel is known in the triangle $\Delta_T$ at the outset,  before the evaluation begins. However,
in the application considered here, the kernel only becomes known as
the time-stepping scheme progresses.  Specifically, at time $t$, 
 $k(t,\tau)$ is only known in the smaller triangle
$(t,\tau) \in \Delta_t$, which is the subset of $\Delta_T$ where $\tau  \leq t$, shown to the left of the blue line in Figure \ref{triangle}.  In the remainder of this
section we explain why the original fast method cannot be used in
this situation and describe a modification to handle the causality of
the kernel.

Fast evaluation methods for integral operators are typically based on two key steps

\begin{enumerate}

\item {\bf Partitioning the domain.} In our fast method, the domain $\Delta_T$ is divided  into diagonal triangles and
  squares that increase in size further away from the diagonal, as
  illustrated in Figure~\ref{triangle}.  Let $P$ denote  the number of time steps along a line $0 \leq \tau \leq t$, and
  let $S(P)$ represent  the number of squares touching the diagonal $t=\tau$, where two triangles are also counted as one square.  
  The step size in $\tau$ is uniform, and when it is halved, i.e., the number of time
  time steps $P$ is doubled with $t$ fixed,  the triangle and square subdomains that touch the
  diagonal are refined  by halving their linear dimensions
  (hence each diagonal subdomain
  contains the same number of time steps throughout the refinement process). This doubles the number of points along the line $t=\tau$ but only refines the partition along the diagonal subdomains.  When the triangle and square subdomains are refined, a new
  ``level" of the partition is created. 
  
  The growth of $S(P)$ during refinement is found to satisfy the recurrence relation
  \[
  S(P)=2S(P/2)+1,
  \]
 so that by the Master Theorem for recurrence relations, 
  $S=O(P)$. A similar argument shows that an additional $O(P)$ squares that do not touch the diagonal are created
 during refinement. 
 Consequently, the total number of squares and triangles in the partition of 
 $\Delta_T$ is $O(P)$.
 
 The number of time steps can also be doubled by doubling the final time $t$ while keeping the time-step size fixed.  The partition
 $\Delta_T$ is duplicated along the line $t=\tau$ and an additional coarse off-diagonal level is created in $\Delta_{2T}$.
  
\item {\bf Kernel approximation via degenerate expansions.}  In regions where the kernel is smooth, it can be approximated using a degenerate kernel expansion.  For Abel integral operators these regions 
  correspond to the squares that do not touch the diagonal.
  
\end{enumerate}

In the $n$-th time step the evaluation of $\K g(t_n)$ contains an
integral over $0\leq \tau \leq t$, which is indicated by the blue
line in Figure~\ref{triangle}. In the discretized version
\eqref{discretize} this is a summation over the node points $t_j$,
$1\leq j \leq n-1$. The cost of evaluating the contributions from the
two endpoints is of lower order and is therefore not considered further.

There are several options for approximating the kernel by a degenerate
series on squares that do not touch the diagonal by an approximation that separates the $t$ and $\tau$
variables. For example, a standard  interpolation at Chebyshev nodes
does not involve derivatives of the kernel, and uses the approximation
\begin{equation}\label{kernel:TwoVar}
  \frac{k(t,\tau)}{\sqrt{t-\tau}} \approx \sum_{k,\ell=1}^q (\omega_k^I-\omega_\ell^J)^{-\half} 
  k(\omega_k^I,\omega_\ell^J) L_k^I(t) L_\ell^J(\tau),\quad
  (t,\tau) \in I \times J.
\end{equation}
Here $I$ and $J$ are the intervals  in the $t$- and $\tau$-axes of an off-diagonal square in the partition of Figure \ref{triangle}.
Further, $\omega_k^I,\omega_\ell^J$ are the Chebyshev interpolation nodes on $I$ and $J$, $L_k^I(\cdot)$ and $L_\ell^J(\cdot)$ are the corresponding
Lagrange polynomials, and   $q$ is the interpolation order that sets
the accuracy of the approximation.    
Since the expansion (\ref{kernel:TwoVar}) is used   on off-diagonal squares the difference
$\omega_k^I-\omega_\ell^J$ is always bounded away from zero and positive. 
 
As a result of (\ref{kernel:TwoVar}),  the sum over Chebyshev nodes
in $I\times J$ has $q^2$ terms. Due to the separability  of (\ref{kernel:TwoVar}) in  $L_k^I(t)$ and $L_l^J(\tau)$,  the matrix-vector product  for integration of (\ref{kernel})  over $\tau$ can
be computed in $O(q^2 P_J)$ operations  instead of the  $O(P_J^2)$ required for a full matrix, where $P_J
= P_I$ is the number of nodes along one direction of the square.
 Since the  subdivision of $\Delta_T$ consists of  $\Ord(P)$ squares and triangles,  the overall computational complexity for the convolution is reduced to $\Ord(q^2 P)$.

However, in the application considered here, the series in
\eqref{kernel:TwoVar} cannot be computed in this manner because at a given time step
$t$ some of the $\omega_k^I$-nodes will be greater than $t$, rendering
the expression $k(\omega_k^I,\omega_\ell^J)$ undefined.

We modify the fast evaluation method as follows. We
only interpolate in the $\tau$-variable
\begin{equation}\label{kernel:OneVar}
\frac{k(t,\tau)}{\sqrt{t-\tau}}  \approx \sum_{\ell=1}^q (t-\omega_\ell^J)^{-\half} 
k(t,\omega_\ell^J) L^J_\ell(\tau),\quad (t,\tau) \in I\times J,
\end{equation}
which ensures that all kernel values are defined at time $t$.

Let $f^I$ be the result of the matrix
vector product corresponding to nodes in $I$ and $J$. Rearranging
terms in  (\ref{kernel:OneVar}) gives
\begin{equation}\label{fast:fi}
    %f^I_i 
    f^I_i = \sum_{\tau_j\in J} (t_i-\tau_j)^{-\half} k(t_i,\tau_j)  g(\tau_j)
    \approx \sum_{\ell=1}^q (t_i-\omega_\ell^J)^{-\half}
    k(t_i,\omega_\ell^J) \mu_\ell(J), 
\end{equation}
for each $i$ such that $t_i \in I$. Here the moments 
\begin{equation} \label{moments}
\mu_\ell(J) = \sum_{\tau_j\in J} L^J_\ell(\tau_j) g(\tau_j)
\end{equation}
are computed for all intervals $J$ with the usual upward pass in $\Ord(P)$
operations.

\underline{\em{Complexity}}.  
As noted, a new level is added when $P$ is doubled, which doubles the number of off-diagonal squares. Hence in level $l$ of  the partition $\Delta_T$  there are $O(2^l)$ off-diagonal squares that each contain $O(2^{-l}P)$ quadrature points $t_i$ in a linear direction, where the level number $l$ satisfies $2 \leq l \leq L$ with $L=O(\log_2 P)$.   The calculation of  $f_i^I$ in (\ref{fast:fi}) for all $t_i \in I$  therefore  has complexity $O(q 2^{-l} P)$ operations  per square in level $l$ with $q$ Chebyshev nodes. This sum must be computed once for each square at a total cost
\begin{equation*}
\sum_{l=2}^L \Ord(q 2^{-l} P) \Ord(2^l) = \Ord( q L P).
\end{equation*}

Since Chebyshev interpolation for smooth functions is exponentially convergent, it can be shown that increasing $q$  so that $q=O(L)=O(\log_2 P)$ ensures that the fast method exhibits the same convergence rate as a nonaccelerated evaluation.  The total complexity is therefore $O(P \log_2^2 P)$ for the off-diagonal evaluation, while the corrected trapezoidal rule quadrature in the diagonal subdomains is $O(P)$.  For more details of the error analysis refer to \cite{tausch2012fast}.

\underline{\em{Modifications for surfactant}}. Another modification needs to be made to the discretization  \eqref{discretize} for the surfactant problem, which solves for $\left. \frac{\partial C}{\partial N} \right|_{N=0}(t_n)$  via  (\ref{kernel equilibrium}). One of the quantities  required to evaluate \eqref{discretize} is $\partial _\tau h_0(t_n)$, which appears in  $r_0\varphi(t_n,t_n)$  where   $\varphi$ is defined in \eqref{varphi}. However, from (\ref{h0}) 
\begin{equation}\label{dth0}
    \partial _\tau h_0(t_n) = \frac{\partial_\tau \Gamma_0(t_n)} {K(1-\Gamma_0(t_n))^2}
\end{equation}
where from (\ref{surfGamN})
\begin{equation}\label{dgamdt}
    \partial_\tau\Gamma_0(t_n) = \left. \Tilde{R}(t_n)+J_0\frac{\partial C}{\partial N} \right|_{N=0}(t_n),
\end{equation}
with $\Tilde{R}(t_n)$  defined in \eqref{R(t)}. Here we can see that the relation between  $\left. \frac{\partial C}{\partial N} \right|_{N=0}(t_n)$  and   $\partial_\tau h_0(t_n)$ 
is implicit, not explicit, but it is linear.

   To handle this, note   
from (\ref{limiting_value}),  \eqref{dth0} and \eqref{dgamdt} that 
\begin{equation}\label{varphi(tn,tn)}
\left.    \varphi(t_n,t_n) = k(t_n,t_n)\partial _\tau h_0(t_n) = \frac{1}{\sqrt{\pi}K(1-\Gamma_0(t_n))^2} \left[ \Tilde{R}(t_n)+J_0\frac{\partial C}{\partial N} \right|_{N=0}(t_n) \right],
\end{equation}
so that  \eqref{discretize} evaluated at time $t_n$ can be rewritten as
\begin{multline} \label{strategy equation}
 \left.   -\frac{\partial C}{\partial N} \right|_{N=0} (t_n) \times \left[1+h^{1/2}r_0\frac{J_0}{\sqrt{\pi}K(1-\Gamma_0(t_n))^2} \right]  \\
= 
h \sum_{j=1}^{n-1} (t_n-t_j)^{-\half} \varphi(t_n,t_j) 
 +  h^\half \Big( r_0 \frac{\Tilde{R}(t_n)}{\sqrt{\pi}K(1-\Gamma_0(t_n))^2} + r_1 \varphi(t_n,t_{n-1}) \Big) \\
 + s_n^{(0)} \varphi_0(t_n) + s_n^{(1)} \varphi_1(t_n)  +  s_n^{(2)} \varphi_2(t_n),
\end{multline} 
which is used to compute $\left. \frac{\partial C}{\partial N} \right|_{N=0}(t_n)$.  
We  calculate  $\partial_\tau h_0(t_n)$ from  \eqref{dth0} and  \eqref{dgamdt},  then store it for the time history, and if needed recompute the moments in \eqref{moments}.

\underline{\em{Solution of the ODE}}.
We now turn to the numerical solution of the ODEs \eqref{psiODE}
which must occur simultaneously with the time convolution, since  $k(t,\tau)$ of (\ref{k(t,tau)}) is defined by the hierarchy of $\psi_0,~ \psi_1$ and $\psi_2$. 

Many methods are available for time
integration of the ODEs.  However,  interpolation points in
the degenerate kernel approximation \eqref{kernel:OneVar} are
typically located between mesh points of the ODE solver. Therefore the ODE solver should
provide a natural way to extend the solution from  mesh points to
 interpolation points.

A suitable integrator is the Adams-Bashforth method. We
recall the derivation in which
$y'(t) = f(t,y)$ is a generic differential equation, or in integral
form  
\begin{equation*}
y(t) = y(t_n) + \int_{t_n}^t\, f(\tau, y(\tau))\, d\tau.
\end{equation*}  
In the Adams-Bashforth method the integrand is replaced by the
interpolate at the previous time steps $t_n,\dots,t_{n-p}$, where $p$ is the order
of the method. This leads to
\begin{equation}\label{AB:between}
Y_n(t) = y_n + \sum_{k=0}^p  f(t_{n-k}, y_{n-k})  \int_{t_n}^t L^p_k(\tau) \ d\tau, 
\quad t \in [t_n, t_{n+1}],
\end{equation}  
where $L^p_k(\cdot)$ is the Lagrange polynomial based on the
previous $p+1$ time steps. Thus $Y_n(t)$ 
is a polynomial approximation of degree $p+1$ for the solution in
the interval $t\in [t_n, t_{n+1}]$, and substituting $t=t_{n+1}$ gives the
next approximation $y_{n+1} = Y_n(t_{n+1})$. The integrals
\begin{equation*}
 a_k = \int_{t_n}^{t_{n+1}} L^p_k(\tau)  \ d\tau
\end{equation*}
are the well-known coefficients of the $p$-th order Adams-Bashforth
method.

In the application considered here the component $k(t,\tau)$ in the kernel of  the integral operator
is defined by \eqref{k(t,tau)}, where $\psi_1(\cdot)$ and $\psi_2(\cdot)$ are
found by the Adams-Bashforth method.  Hence, when interpolating the
kernel in \eqref{kernel:OneVar} the polynomials \eqref{AB:between}
approximate  $\psi_2(\cdot)$ at the Chebyshev
interpolation nodes.
Since the accuracy of the fast convolution method is limited by
  the number of terms in \eqref{h0ste}, it suffices to use
  $p=2$ in \eqref{AB:between}.  As noted, a second-order Adams-Bashforth scheme is used to evolve  $s_\alpha$, $\theta$, and $\Gamma$  in (\ref{salpha}), (\ref{theta}),  and (\ref{surfGamN}).  As implemented, the method achieves spectral accuracy in space and $O(h^{3/2})$ accuracy in time.
  
  {\underline{\em Bulk surfactant concentration.}}  The bulk surfactant concentration is found by evaluating (\ref{Bulk C Solution}) during a post-processing step. Special care is needed to handle the singularity in the integrand at $u = 0$. Following the approach in \cite{xu2013analytical}, the integral is split into two parts: in $C_1$,  $u \in [0, \delta]$, where $\delta > 0$ is small;  in $C_2$,  $u \in [\delta, t]$. Then, \begin{equation}\label{C1 integral} C_1(N, t) = \frac{2h_0(t)}{\sqrt{\pi}} e^{\frac{\psi_0(t)N^2}{4}} \int_{\frac{N}{2\sqrt{\delta}}}^{\infty} e^{-s^2} \left(1 + O\left(\frac{N^2}{s^2}\right)\right) ds,
 \end{equation} where $s = N / (2\sqrt{u})$, which gives
 \begin{equation}\label{C1 exact} C_1(N, t) = \frac{2h_0(t)}{\sqrt{\pi}} e^{\frac{\psi_0(t)N^2}{4}}  \text{erfc}\left(\frac{N}{2\sqrt{\delta}}\right) + O(\delta). 
  \end{equation}
Note that as $N \to 0$, the integral in (\ref{C1 integral})  tends to $\sqrt{\pi}/2$ and $C_1$ approaches $h_0(t)$, so that from (\ref{Bulk C Solution})  $C(s_0, N = 0, t) = 1 + h_0(t)$, recovering the boundary condition from (\ref{Ceqn_init_BC}).

The expression for $C_2$ is given by 
\begin{equation}\label{C2 integral} C_2(N, t) = \frac{2N}{\sqrt{\pi}} \int_{\delta}^{t} e^{-(N/\gamma)^2} \frac{\partial_u \gamma}{\gamma^2} h_0(t - u) \ du. \end{equation}
In post-processing, $C_1$ is computed from  (\ref{C1 exact}) with a typical value of $\delta = 0.01$, while $C_2$ is evaluated from  (\ref{C2 integral}) by the trapezoidal rule. The total bulk concentration is then given by $C = 1 + C_1 + C_2$.  This procedure yields an  $O(\delta)$ error, which is acceptable for plotting data.

The discretization and fast evaluation method, along with its order of accuracy, are validated in the next section. Further validation using a synthetic example is presented in
Appendix A.

\subsection{Mesh-Based Method}\label{Mesh}

We compare the fast mesh-free method to a mesh-based method developed in 
 \cite{xu2013analytical}, which  solves (\ref{bulkCMesh})  to  (\ref{BCMesh})  using an equal-arc--length parameterization.   In this approach, the arc length satisfies  $s_\alpha(\alpha,t)=l(t)$, so that $\p_\alpha s$  is independent of  $\alpha$ but depends on time.  The parameterization is enforced by setting the tangential velocity $\phi_s$ to satisfy (\ref{phi_s_equal_arclength}). The spatial domain for $C(\alpha,N,t)$ is a rectangular region $ \alpha \times N \in \mathcal{R}=[0,2\pi] \times [0, N_{max}]$,  truncated at $N=N_{max}$.  On this   domain the far-field boundary condition in the second equation of  (\ref{BCMesh0})  is replaced by
 \be \label{truncated_far-field}
 C(\alpha,N_{max},t)=1~~\mbox{for all}~~t>0.
 \ee
 A more refined far-field boundary condition that accounts for inflow-outflow effects at $(\alpha, N_{max})$ is given in \cite{wang2014numerical}.

Chebyshev-Lobatto collocation points are introduced in the $N$-direction of $\mathcal{R}$ and a Chebyshev differentiation matrix is used to discretize all  $N$-derivatives in  \eqref{bulkCMesh}.  The surface parameter $\alpha$ is discretized with a uniform step size  $\alpha_j  \in [0, 2 \pi)$ for $j=1, \ldots, N_s$, and  $\alpha$-derivatives are computed using the FFT. 

This method achieves spectral accuracy  in space and is implemented with first-order accuracy in time.  
 In principle, the method can be easily extended to be higher order in time, but first order suffices for our comparison with the fast mesh-free method.  Further details on the mesh-based method can be found in \cite{xu2013analytical}.

The complexity of the mesh-based method is $O(M^2P)$ operations per 
$\alpha_j$ surface mesh point, where $M$ represents the number of mesh points in the normal or $N$-direction within the transition layer. In some instances, the size of the mesh $M$ can increase over time, significantly raising the computational cost.

\captionsetup[table]{singlelinecheck=false}

\section{Validation}\label{OoA}

\subsection{Order of accuracy}

\begin{figure}[h]
        \centering
        \includegraphics[scale=.33]{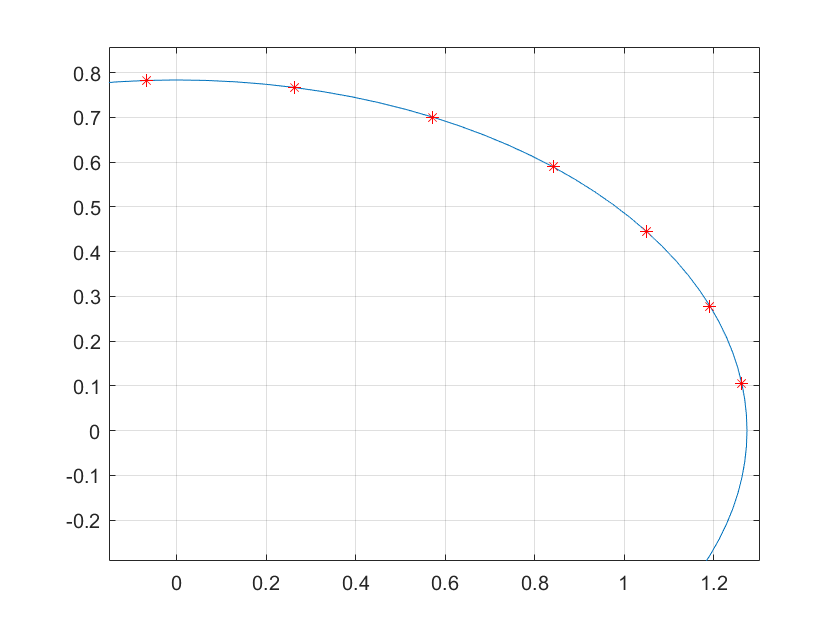}
         \includegraphics[scale=.33]{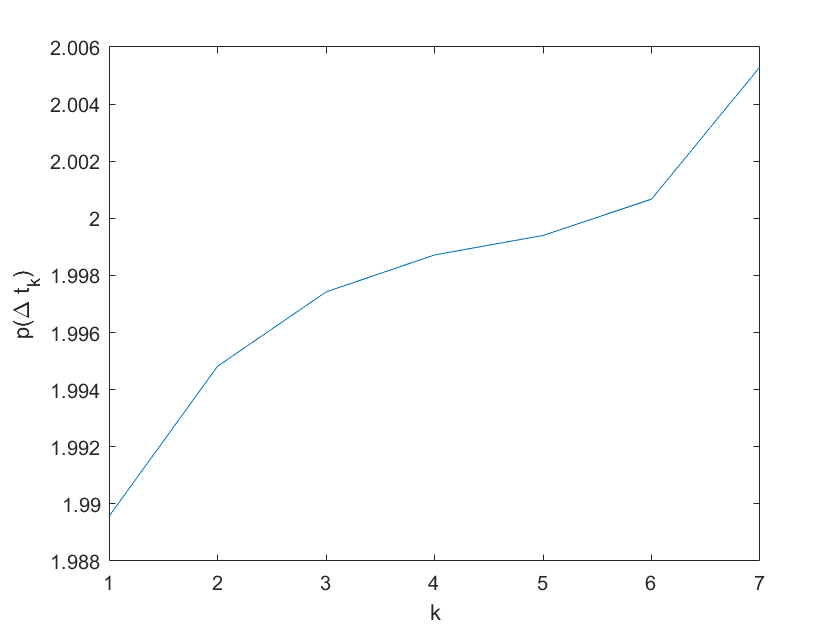}
        \caption{(a) Location of equispaced points in $\alpha$ for the data in Table \ref{Ooa dcdn table exact A2} at time $t=1.28$.  (b) Order of accuracy $p(\Delta t_k)$ for the area of the drop.}
        \label{ooa graph area}
\end{figure}

The order of accuracy $p$ of the fast mesh-free method to compute a quantity $f$  is assessed using the formula
\begin{equation}\label{OoA2 eq4}
    p(\Delta t) =  \log_2 \frac{|f_{\Delta t}-f_{\Delta t/2}|}{|f_{\Delta t/2}-f_{\Delta t/4}|},
\end{equation}
where $f_{\Delta t}$ is computed with time step $\Delta t$ and sufficiently high spatial resolution to ensure that the dominant source of error comes from the time discretization.   We present detailed results for $p$ in Table \ref{Ooa dcdn table exact A2} when  $f$ is $\left. \frac{\partial C}{\partial N} \right|_{N=0}(\alpha)$. These are evaluated at equispaced values of the Lagrangian marker $\alpha$ over one quarter of the symmetric drop shape.
The position of the markers at $t=1.28$ is shown in Figure \ref{ooa graph area}a.

 For the data presented in the table, we consider an initially circular drop in a pure strain flow with parameters $Q=0.5$, $\lambda=0.01$, $J_0=1$, $E=0.1$, $K=1$, $\Gamma(\alpha,t=0)=0.5$, and $N_s=128$ interface points.  Expansion coefficients $A_0,~A_1$ and $A_2$ are computed from (\ref{A0A1}) and (\ref{A2 surf}).   The initial time step is set to   $\Delta t = 0.005$,  and we define $\Delta t_k = \Delta t/2^k$ for $k = 1, 2, \ldots$ to compute the order of accuracy $p(\Delta t_k)$  for $k=1,2,...,6$.
 The results show good agreement with the expected $O(h^{3/2})$ accuracy of the method. Similar results are obtained for other quantities such as $s_\alpha,  \theta$, and $\Gamma$.

\begin{table}[h]
        \caption[Order of Accuracy for $\frac{\partial C}{\partial N}|_{N=0}$ with Exact Value of $A_2$]{Order of accuracy for $\frac{\partial C}{\partial N}|_{N=0}$ at a set of interface points.}
    \centering
    \setlength\tabcolsep{4pt}
    \resizebox{\textwidth}{!}{\begin{tabular}{||c||c|c|c|c|c|c|c||} \hline  
        k & $\alpha=4.6633$ & $\alpha=4.9087$ & $\alpha=5.1542$ & $\alpha=5.3996$ & $\alpha=5.6450$ & $\alpha=5.8905$ & $\alpha=6.1359$ \\ \hline  \hline
        1 & 1.2730 & 1.2769 & 1.2981 & 1.4494 & 1.3473 & 1.1541 & 1.1880 \\
        2 & 1.3584 & 1.3602 & 1.3706 & 1.4541 & 1.3934 & 1.3004 & 1.3148 \\
        3 & 1.4057 & 1.4067 & 1.4118 & 1.4573 & 1.4150 & 1.3781 & 1.3839 \\
        4 & 1.3995 & 1.3998 & 1.3945 & 1.3694 & 1.2485 & 1.4809 & 1.4546 \\
        5 & 1.5073 & 1.4967 & 1.5013 & 1.5739 & 1.6347 & 1.3739 & 1.4085 \\
        6 & 1.4565 & 1.4600 & 1.4600 & 1.1457 & 1.5385 & 1.4909 & 1.4821 \\
        \hline
    \end{tabular}}
     \label{Ooa dcdn table exact A2}   
\end{table}

Figure \ref{ooa graph area}b shows accuracy data for the area of the drop, which is a conserved quantity. This is computed using a two-point variant of (\ref{OoA2 eq4}), since the exact value of the area is known.
The order of accuracy for this integrated quantity is found to be
$p=2$.

Extensive tests of the fast mesh-free algorithm for computing the convolution  (\ref{kernel})  using synthetic data are provided in Appendix A.

\captionsetup[subfigure]{font={bf,small},skip=1pt, margin=-0.01cm, singlelinecheck=false}

\subsection{Comparison of Mesh-Based and Mesh-Free Methods}\label{S5 comp}

Here we compare the accuracy and efficiency of the fast mesh-free method of Section \ref{FastHM} with the mesh-based method of Section \ref{Mesh}.  The mesh-based method has been validated previously in \cite{booty2010hybrid} and \cite{xu2013analytical}. In \cite{booty2010hybrid}, the comparison is with a traditional approach that employs an adaptive interface-fitted mesh in the fluid region without the transition layer reduction. Given this prior validation, the mesh-based method serves as a good benchmark for assessing the performance of the mesh-free method.

The fast mesh-free method uses a Lagrangian coordinate frame while the mesh-based method uses an equal-arc--length frame, so to compare the results we need to transform between the two frames. Let $\mathbf{X}(\alpha,t)$ denote the equal-arc--length parameterization of the interface at time $t$, and $\mathbf{X}(\alpha,0)=\mathbf{X}(\alpha_0)$ represent the arc-length parameterization of the intial interface shape. 
The location of the same Cartesian point  in the Lagrangian frame  is given by   $\mathbf{X}(\alpha_m(\alpha,t),t)$
where the {\em forward} map $\alpha_m(\alpha,t)$ is such that $\alpha_m$  gives the location of the material point that at time $t$ is located at arc-length parameter $\alpha$, i.e., $\mathbf{X}(\alpha_m(\alpha,t),t)=\mathbf{X}(\alpha,t)$.
The  forward map satisfies \cite{higley2012semi}
\begin{equation}\label{formap}
    \frac{\partial\alpha_m(\alpha,t)}{\partial t} = \frac{u_s(\alpha_m(\alpha,t),t)-\phi_s(\alpha_m(\alpha,t),t)}{s_\alpha(t)}
\end{equation}
with initial condition $\alpha_m(\alpha,0)=\alpha_0$.
The forward map is only used during the post-processing of data to transform between the two parameterization frames.

The fast mesh-free method has been validated against the mesh-based method by comparing the evolution of drop profiles,  interfacial surfactant concentration $\Gamma$, the normal derivative of the bulk surfactant concentration at the interface $\frac{\partial C}{\partial N}|_{N=0}$, and the bulk surfactant concentration $C$.  The comparisons were made for a drop subjected to imposed strain and shear flows.

\begin{figure}[h]
\begin{subfigure}{.5\textwidth}
        \subcaption{}
        \centering
        \includegraphics[scale=.3]{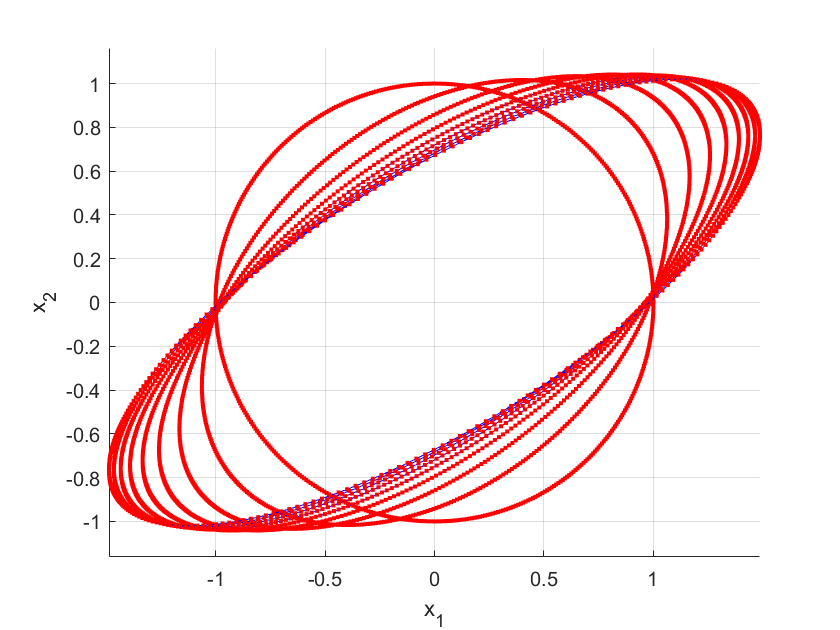}
        \label{profile shear}
    \end{subfigure}
    \begin{subfigure}{.5\textwidth}
        \subcaption{}
        \centering
        \includegraphics[scale=.3]{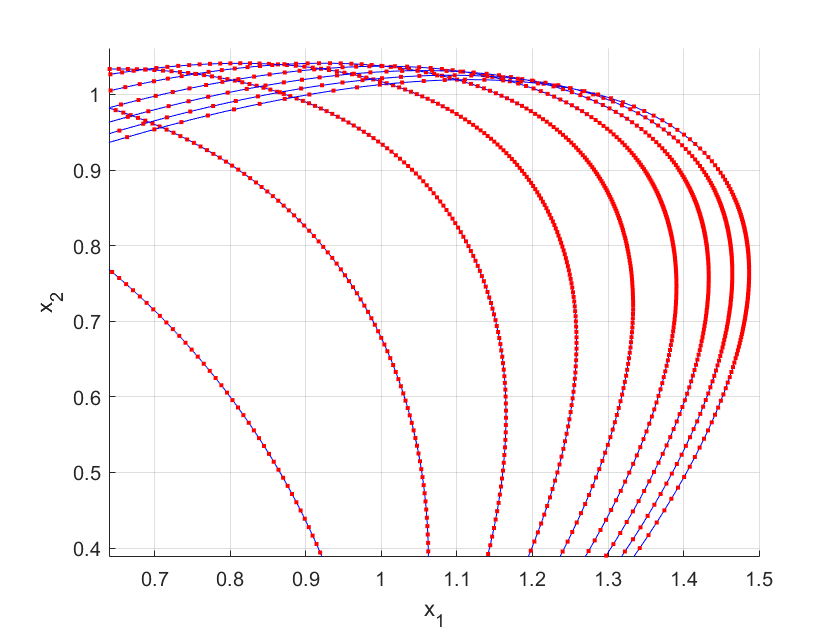}
        \label{profile shear zoom}
    \end{subfigure}
    \caption[Comparison of drop profiles of a drop stretched in a shear flow with $G=.5$]{Comparison of mesh-based and fast mesh-free methods for a drop stretched in a shear flow. Data is indicated by red dots for the mesh-free method, and by blue lines for the mesh-based method. (a) Drop profiles from  $t=0$ to $t=4.096$ plotted at time interval $\Delta t_p=0.512$. (b) Close-up view near a drop tip.}
    \label{Shear profile Acc}
\end{figure}

As an example, consider an intially circular drop deformed by a shear flow with parameters $G=2B=0.5$, $Q=0$, viscosity ratio $\lambda=0.01$, and surfactant parameters  $K=1.5$, $J_0=1$, and  $E=0.1$, so that the initial uniform surface surfactant concentration is  $\Gamma(\alpha,t=0) = 0.6$, ensuring the bulk and interfacial surfactant concentrations are initially in equilibrium, see equations (\ref{BCbulkC}) and (\ref{ICbulkC}).  Figure \ref{profile shear} compares interface profiles obtained from the two methods as the drop evolves from $t=0$ to $t=4.096$ with time step   $\Delta t=5\times 10^{-4}$.  The interval between recorded drop profiles is   
$\Delta t_p=0.512$.  The interface is discretized with  $N_s=512$ points, and the mesh-based method has $M=256$ points in the direction normal to the interface with cutoff at $N_{\text{max}}=40$.  
The results demonstrate excellent agreement between the two methods, as is illustrated further in Figure \ref{profile shear zoom} which shows a close-up view of one of the drop tips.

\begin{figure}[h]
\begin{subfigure}{.5\textwidth}
        \subcaption{}
        \centering
        \includegraphics[scale=.3]{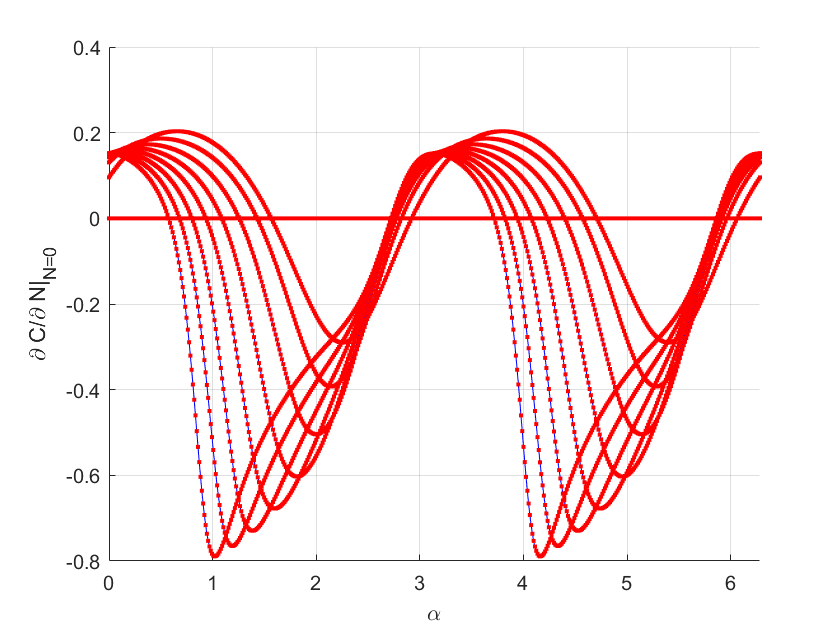}
        \label{dcdn shear}
    \end{subfigure}
    \begin{subfigure}{.5\textwidth}
        \subcaption{}
        \centering
        \includegraphics[scale=.3]{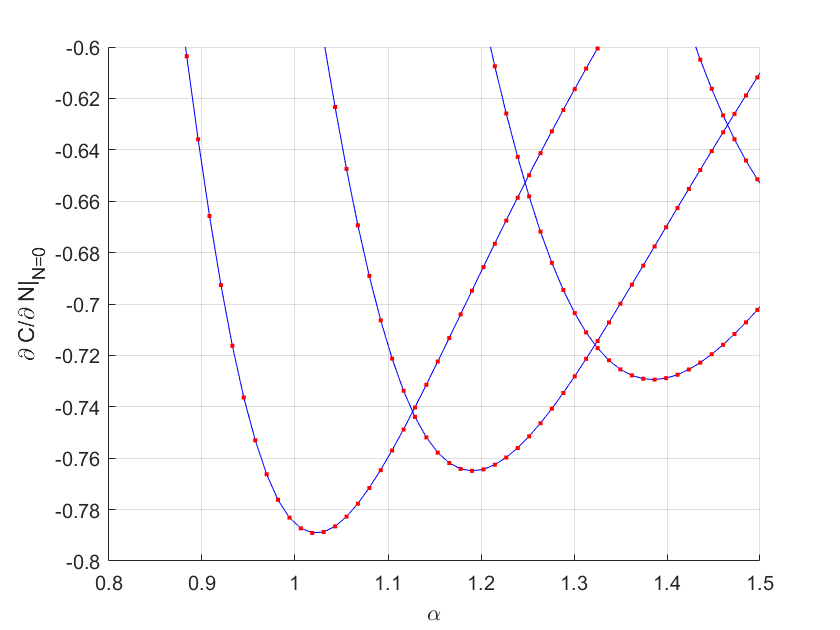}
        \label{dcdn shear zoom}
    \end{subfigure}
    \caption[Comparison of bulk interface surfactant exchange term,  of a drop stretched in a shear flow with $G=.5$]{(a) Comparison of the bulk-interface surfactant exchange term $\frac{\partial C}{\partial N}|_{N=0}$  from the mesh-based and fast mesh-free methods for the simulation of Figure \ref{Shear profile Acc}. (b) Close-up near a drop tip. The horizontal axis $\alpha \in [0,2 \pi)$ is a Lagrangian parameter.}
    \label{shear dcdn Acc}
\end{figure}

Figure \ref{shear dcdn Acc} shows comparison of the bulk-interface surfactant exchange term $\frac{\partial C}{\partial N}|_{N=0}$
computed by both methods at the same times as in Figure \ref{Shear profile Acc}.
With an imposed shear flow, interfacial surfactant accumulates at the drop tips,  causing steep negative normal concentration gradients in $C$, see Figure \ref{shear dcdn Acc}a.  Close-up of the data near a drop tip is shown in  Figure \ref{shear dcdn Acc}b. 

\begin{figure}[h]
\begin{subfigure}{.5\textwidth}
        \subcaption{}
        \centering
        \includegraphics[scale=.3]{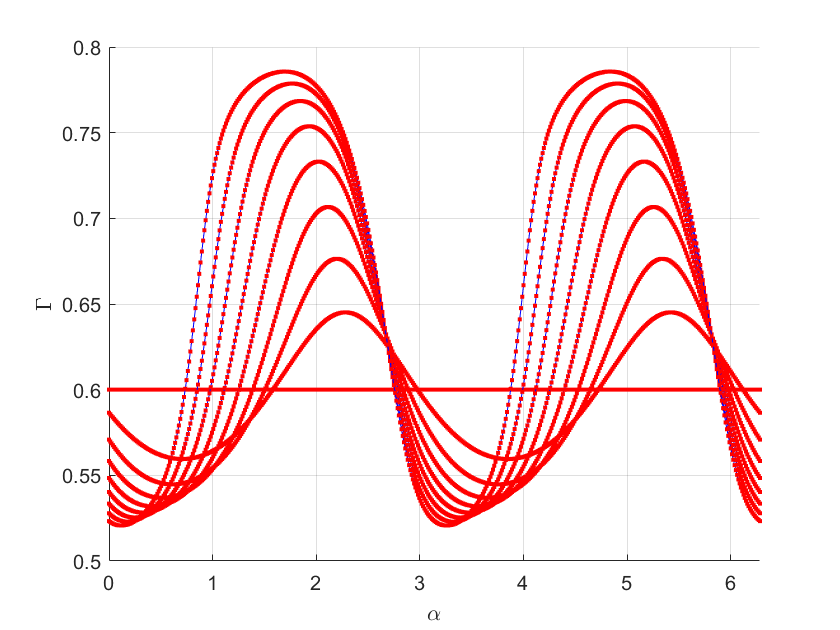}
        \label{gam shear}
    \end{subfigure}
    \begin{subfigure}{.5\textwidth}
        \subcaption{}
        \centering
        \includegraphics[scale=.3]{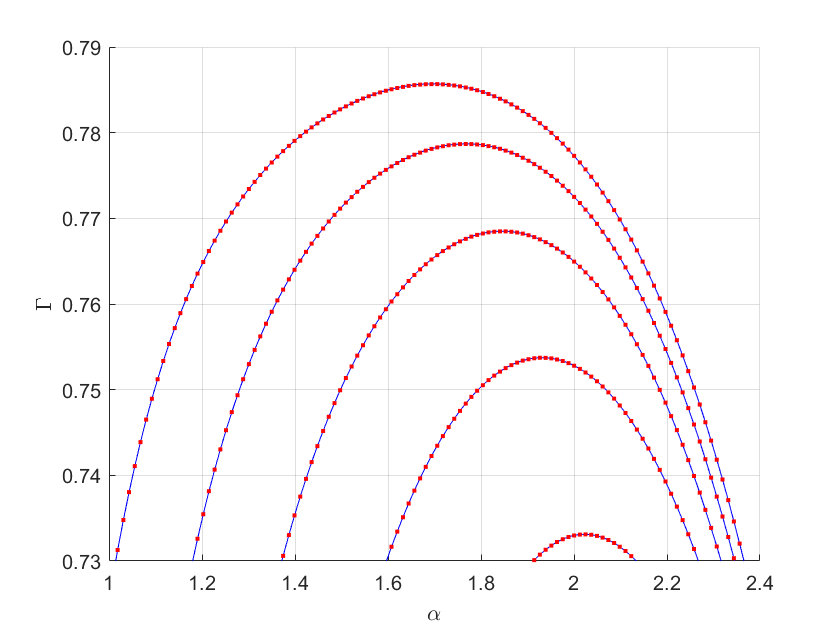}
        \label{gam shear zoom}
    \end{subfigure}
    \caption[Comparison of surface surfactant concentration of a drop stretched in a shear flow with $G=.5$]{(a) Comparison of the interfacial surfactant concentration $\Gamma$  from the mesh-based and fast mesh-free methods for the simulation of Figure \ref{Shear profile Acc}. (b) Close-up near a drop tip. }
    \label{shear Gam Acc}
\end{figure}

Figure \ref{shear Gam Acc} shows the surface surfactant concentration $\Gamma$ at the same times as in the previous two figures.  As noted, $\Gamma$ attains a maximum at the drop tips.  Since $\Gamma$  is highly sensitive to errors in the bulk-interface surfactant exchange term, it serves as a stringent benchmark for evaluating the accuracy of the two methods. The results again show close agreement between the mesh-based and fast mesh-free computations.

\begin{figure}[h!]
\begin{subfigure}{.5\textwidth}
        \subcaption{}
        \centering
        \includegraphics[scale=.32]{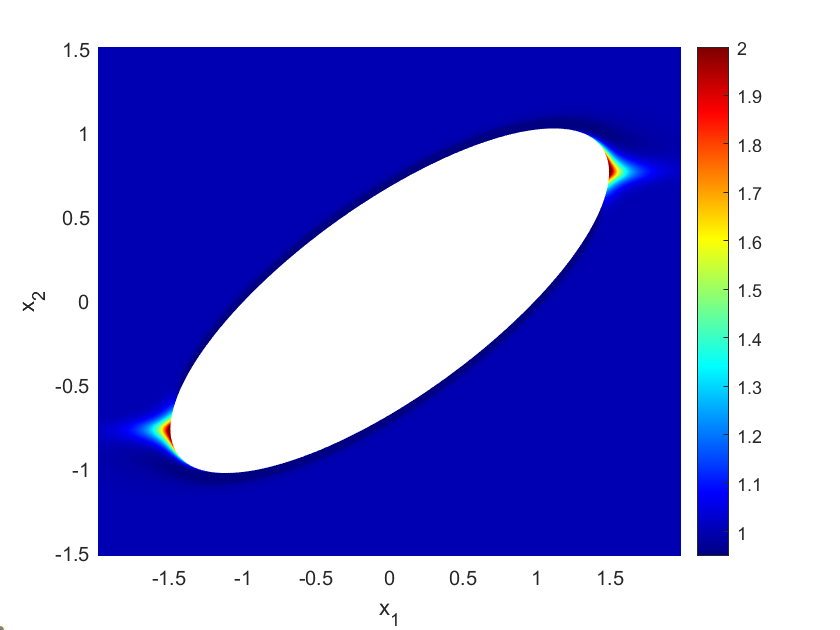}
        \label{bulk shear mesh}
    \end{subfigure}
    \begin{subfigure}{.5\textwidth}
        \subcaption{}
        \centering
        \includegraphics[scale=.32]{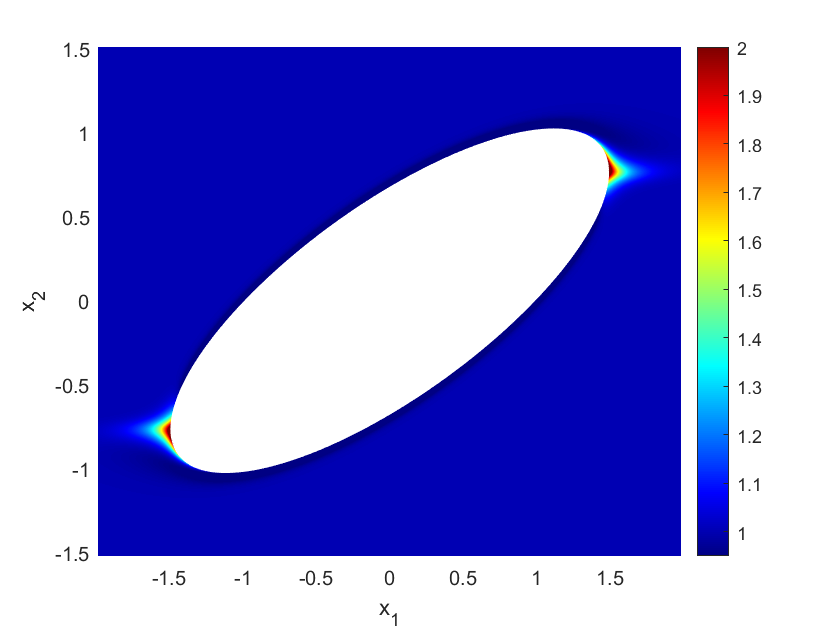}
        \label{bulk shear free}
    \end{subfigure}
    \begin{subfigure}{.5\textwidth}
        \subcaption{}
        \centering
        \includegraphics[scale=.32]{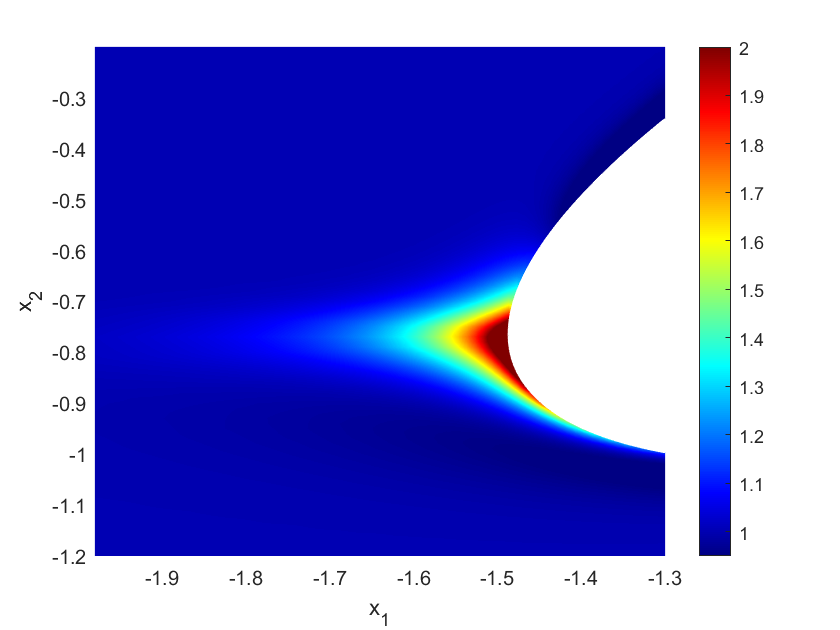}
        \label{bulk shear mesh zoom}
    \end{subfigure}
    \begin{subfigure}{.5\textwidth}
        \subcaption{}
        \centering
        \includegraphics[scale=.32]{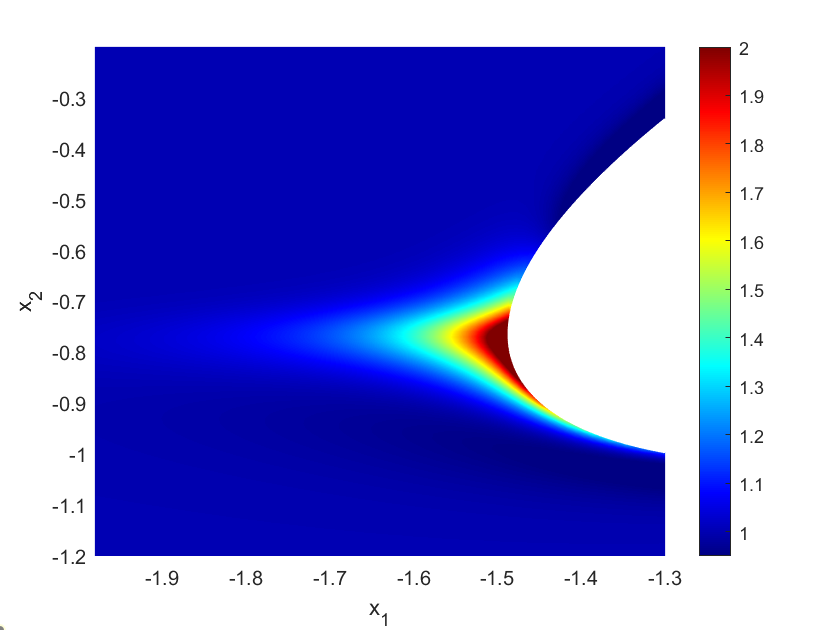}
        \label{bulk shear free zoom}
    \end{subfigure}
    \caption[Comparison of the bulk surfactant concentration of a drop stretched in a shear flow with $G=.5$]{Comparison of the bulk surfactant concentration $C$  from the mesh-based and fast mesh-free methods for the simulation of Figure \ref{Shear profile Acc}. Data is shown at the final time $t=4.096$ with $Pe=3000$. (a) Mesh-based method. (b) Fast mesh-free method. (c)   Close-up of (a) near a drop tip. (d) Close-up of (b) near a drop tip.}
    \label{Shear bulk Acc}
\end{figure}

The bulk surfactant concentration $C$  is computed in-line by the mesh-based method and off-line (i.e., in post-processing) by the fast mesh-free method via (\ref{C1 exact}) and  (\ref{C2 integral}). 
Contraction of the interface near the drop tips causes large surface concentrations and large desorption from the interface into the bulk. Both the surface and bulk concentrations reach a maximum there. Conversely, expansion of the interface occurs where the interface is more flat, causing low surface surfactant concentration and some adsorption from the bulk.  This is seen in the data of Figure  \ref{Shear bulk Acc}.

\subsection{Efficiency}\label{S5 speed comp}

\begin{figure}[h!]
    \centering
    \includegraphics[scale = .5]{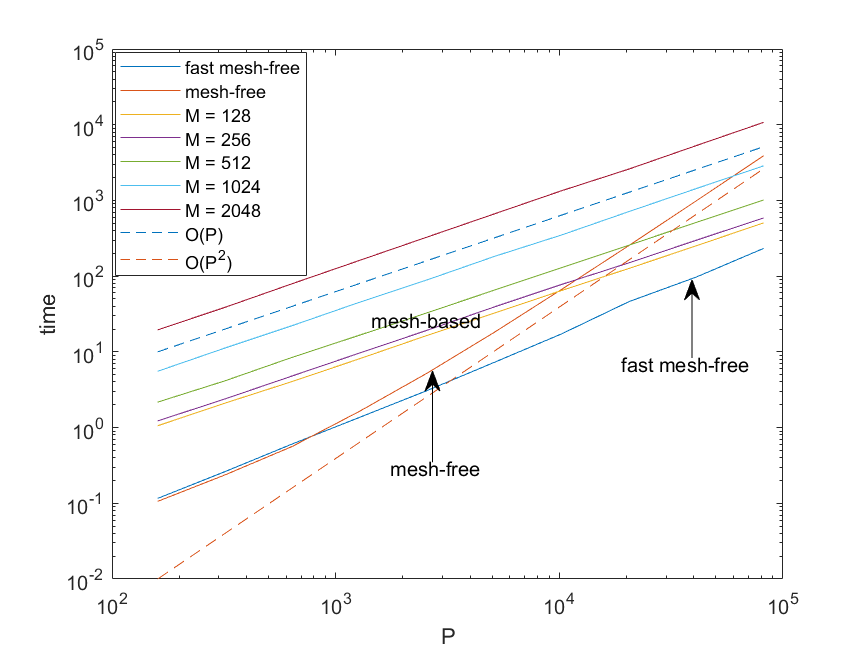}
    \caption[Graphical Representation of Table \ref{Surf Time Table}]{Wall-clock time comparison for computing the bulk-surfactant exchange term versus number of time steps 
$P$. Results for the mesh-based method are shown for various values of 
 $M$. For reference, the dashed lines indicate  $O(P)$ and $O(P^2)$
 scalings.}
    \label{Surf Time Graph}
\end{figure}

Figure \ref{Surf Time Graph}   shows comparison of the wall-clock time required to compute the bulk surfactant exchange term  $\frac{\partial C}{\partial N}|_{N=0}$ using three methods: the fast mesh-free method, the unaccelerated mesh-free method, and the mesh-based method with varying values of the number of mesh points
$M$  in the normal direction that are used to discretize the transition layer. The simulations were performed with a fixed number of interface points 
 $N_s=128$ and final time $t=1.6$. Other parameters for this strain flow are fixed at $G=0$, $B=0$, $Q=0.5$, $\lambda=1$, $J_0=1$, $E=0.1$ and $K=1$. 

The three methods have the following theoretical complexity per interface point:  (i)  $O(M^2P)$ for the mesh-based method, (ii)  $O(P\log_2^2 P)$ for the fast mesh-free method, and  (iii)  $O(P^2)$  for the unaccelerated mesh-free method, where $P$ is the number of time steps.  The numerical results align with these theoretical scalings, particularly in terms of the dependence on 
$P$. Although the mesh-based method has slightly better asymptotic complexity in $P$ than the fast mesh-free method
 (by a factor of $1/\log_2^2 P$),  the fast mesh-free method is observed to be $5–10$ times faster in these representative computations. This speed advantage is due to  the relatively  large values of 
$M$ (typically $256–1024$) required to resolve the transition layer, which, while independent of the P\'eclet number 
$Pe$, substantially increases the computational cost of the mesh-based method and offsets its nominal scaling benefit.

\captionsetup[subfigure]{font={bf,small}, skip=1pt, margin=-0.01cm, singlelinecheck=false}

\section{Numerical examples} \label{sec:NumEx}

Here we present numerical examples to demonstrate the capabilities of the fast hybrid method, which couples the boundary integral fluid solver with the fast mesh-free method for computing the bulk-interface exchange term. To support long-time simulations, we have implemented an adaptive grid strategy that dynamically adjusts the number of interface points during a simulation.

Increasing the number of interface grid points is necessary when a drop is highly stretched, which can cause either under-resolution or clustering of Lagrangian markers.
Conversely, reducing the number of interface points is useful in simulations of drop relaxation, as in Section \ref{swiss}.  
Although reducing the number of points is straightforward, increasing the grid resolution presents a challenge in our fast hybrid method, since it requires a compressed time history of the terms  $\partial_\tau h_0(\tau)$ and $\gamma(t-\tau,\tau)$ from equation \eqref{dcdn}, which are not readily available at new grid points.

We use trigonometric interpolation to reconstruct the necessary time history and then evaluate the moments (\ref{moments}) at the new spatial grid points. 
For efficiency we implement the trigonometric interpolation using the Nonuniform Fast Fourier Transform (NUFFT), which allows rapid evaluation of the interpolants at nonuniformly spaced points with near-optimal computational complexity. This approach ensures that the overhead introduced by adaptive refinement is minimal.

\subsection{Polar rose shapes}

We consider polar roses as a family of initial shapes, where the interface is  parameterized by 
\begin{equation}\label{bunny eq}
    \begin{aligned}
        x_1(\alpha) &= (\sin(r_1\alpha)+\cos(r_2\alpha)+r_3)\cos \alpha \\
        x_2(\alpha) &= (\sin(r_1\alpha)+\cos(r_2\alpha)+r_4)\sin \alpha
    \end{aligned}
\end{equation}
for $\alpha\in[0,2\pi)$.  We give results for two sets  of  $\{ r_i \}$.

\begin{figure}[h]
\begin{subfigure}{.5\textwidth}
        \subcaption{}
        \centering
        \includegraphics[scale=.32]{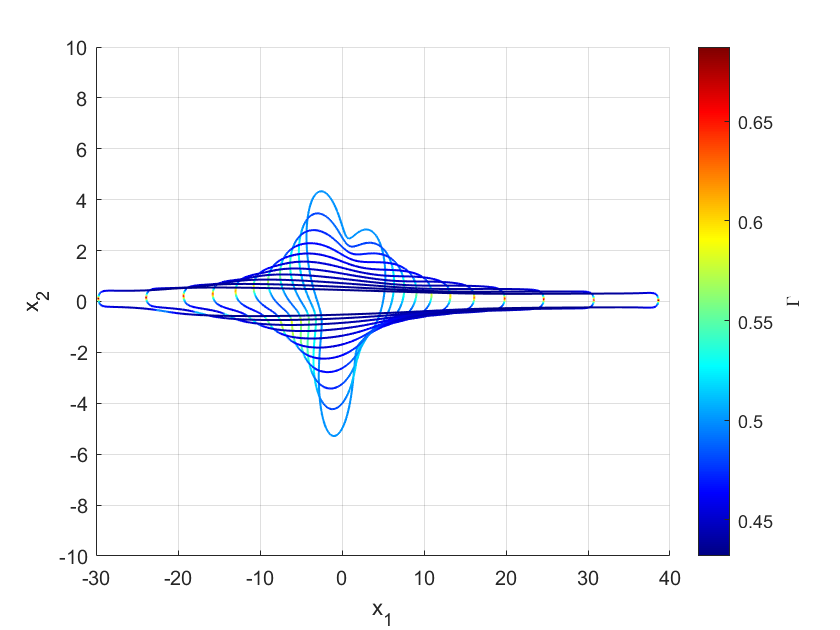}
        
        \label{pr1strain}
    \end{subfigure}
\begin{subfigure}{.5\textwidth}
        \subcaption{}
        \centering
        \includegraphics[scale=.32]{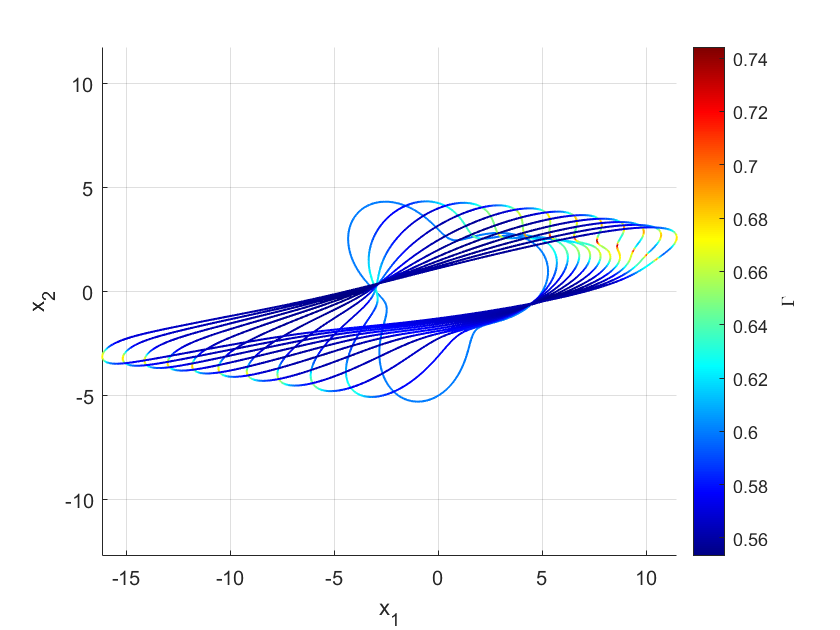}
        
        \label{pr1shear}
    \end{subfigure}
    \caption[Deformation of drop from Figure \ref{PR1 Initial}]{Deformation of an initial polar rose shape given by equation (\ref{bunny eq})  with $r_1 = 3$, $r_2 = 3$, $r_3 = 4$ and $r_4 = 4$ in (a) a strain flow and (b) a shear flow. Note that the axes in (a) are not to scale. }
    \label{PR1Evo}
\end{figure}

The first set is given by   $r_1 = 3$, $r_2 = 3$, $r_3 = 4$, and $r_4 = 4$. The initial shape and its  time evolution under shear and strain flows are shown in Figure \ref{PR1Evo}.  Both simulations used $N_s = 512$ interface points, time step $\Delta t = 5 \times 10^{-4}$, viscosity ratio $\lambda = 1.5$, $J_0 = 1$, and $E=0.1$.  For the strain flow, $Q = 0.5$, $K = 1$, and  the initial equilibrium surfactant distribution is $\Gamma(\alpha, 0) = 0.5$, while for the shear flow, $G = 0.5$, $K = 1.5$, and $\Gamma(\alpha, 0) = 0.6$.
At $t = 2.56$, the number of interface points was doubled to $N_s = 1024$ in both flows to maintain resolution.

Figure \ref{PR1Evo} illustrates the drop deformation for this test case. Figure \ref{PR1Evo}a shows the drop evolution under strain flow, with drop profiles shown from $t = 0$ to $t = 10.24$ at time intervals of $\Delta t_p = 1.024$. Figure \ref{PR1Evo}b shows profiles in a shear flow at the same times.  The surface concentration of surfactant $\Gamma$ is superimposed on the drop profiles.
In both types of flow, the drop deforms into an ellipse-like shape. 

\begin{figure}[h]
\begin{subfigure}{1\textwidth}
        \subcaption{}
        \centering
        \includegraphics[scale=.3]{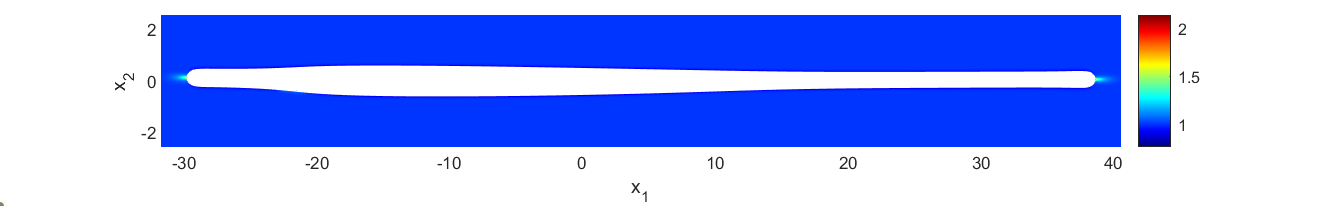}
        
        \label{pr1bulkstrain}
    \end{subfigure}
    \begin{subfigure}{.5\textwidth}
        \subcaption{}
        \centering\
          \includegraphics[scale=.32]{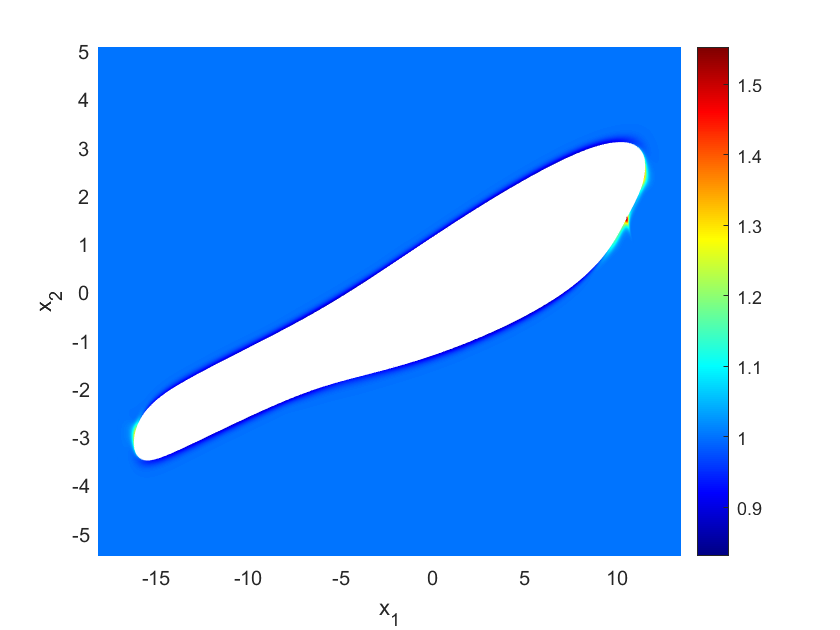}
        
        \label{pr1bulkstrainzoom}
    \end{subfigure}
    \begin{subfigure}{.5\textwidth}
        \subcaption{}
        \centering
                \includegraphics[scale=.32]{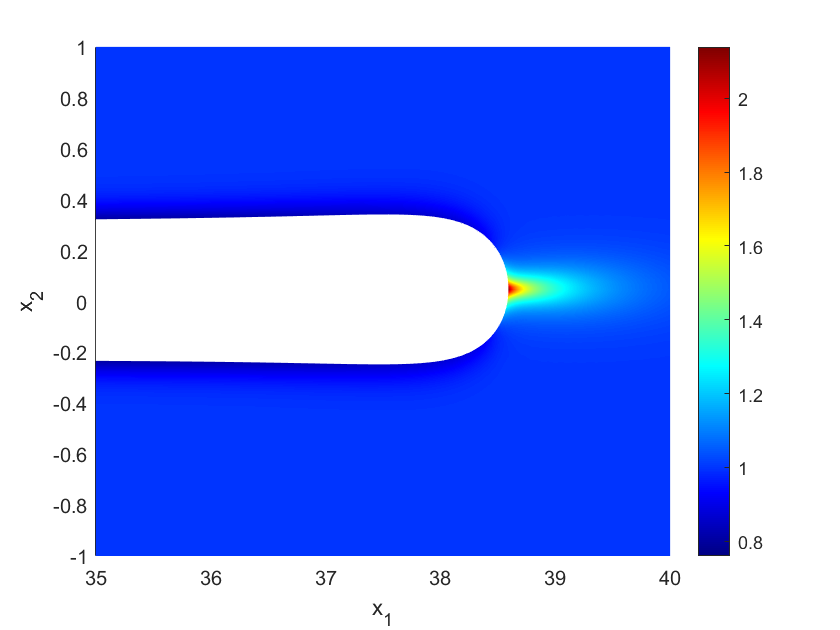} 
        \label{pr1bulkshear}
    \end{subfigure}
    \caption[Bulk concentration of surfactant of drop from Figure \ref{PR1 Initial} at time $t=10.24$]{Bulk surfactant concentration corresponding to the simulations in Figure \ref{PR1Evo} at time $t = 10.24$: (a) Strain flow, (b) shear flow, and (c) close-up of (a) near the drop tip.}
    \label{PR1Bulk}
\end{figure}

Figure \ref{PR1Bulk} shows the bulk concentration of surfactant for the simulations  of Figure \ref{PR1Evo} at the final time $t = 10.24$. In the post-processing step to compute $C$ via 
\eqref{C1 exact} and \eqref{C2 integral}, $M = 128$ points were used in the normal direction, with $\delta = 0.01$, and $Pe = 1000$. 
In both examples, surfactant desorbs into the bulk near the drop tips, where the surface concentration of surfactant is greatest.

\begin{figure}[h]
\begin{subfigure}{.5\textwidth}
    \subcaption{}
        \centering
        \includegraphics[scale=.32]{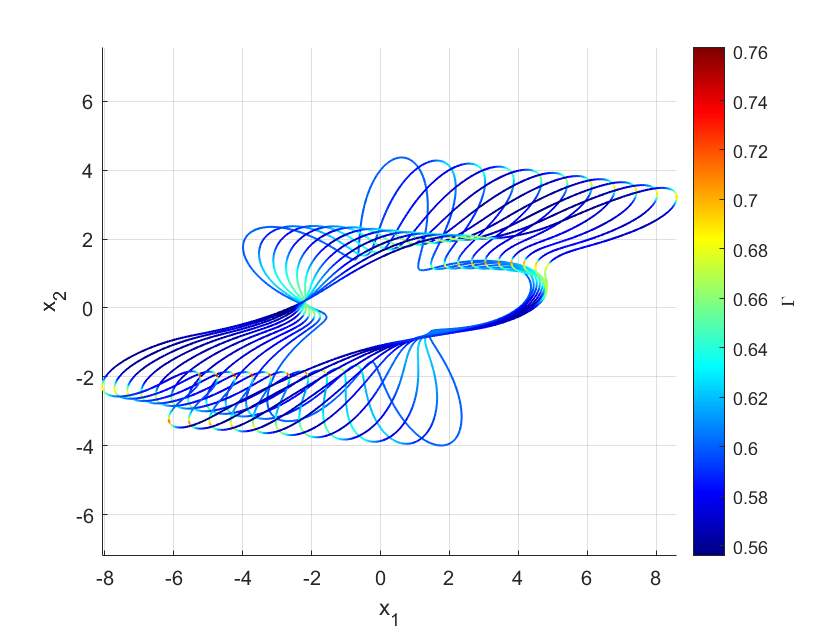}
        
        \label{pr2shear}
    \end{subfigure}
    \begin{subfigure}{.5\textwidth}
    \subcaption{}
        \centering
        \includegraphics[scale=.32]{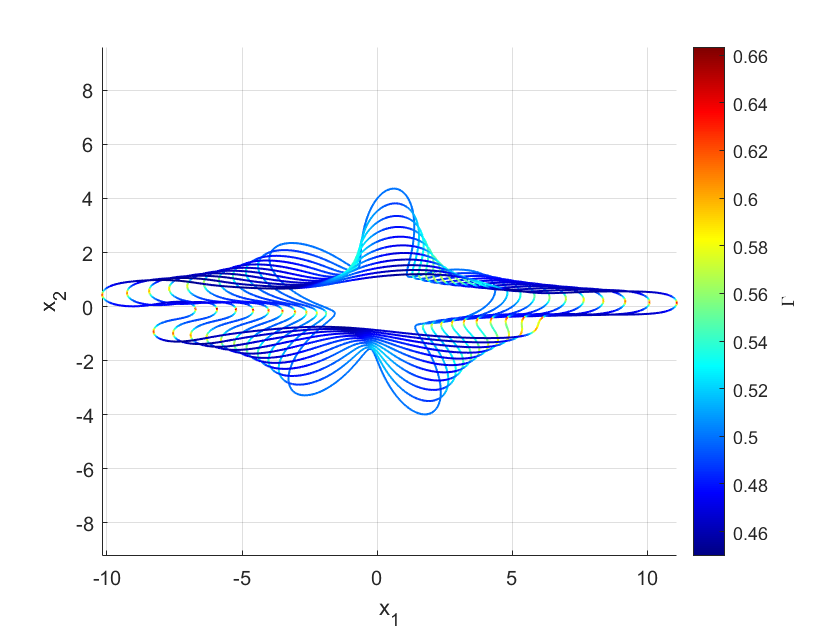}
        
        \label{pr2strain}
    \end{subfigure}
    \caption[Deformation of drop from Figure \ref{PR2 Initial}]{Deformation of an initial polar rose shape given by equation (\ref{bunny eq})  with  $r_1 = 5, r_2 = 5, r_3 = 3, $ and $r_4 = 3$ in (a) a strain flow and (b) a shear flow. }
    \label{PR2Evo}
\end{figure}

In the second example,   $r_1 = 5, r_2 = 5, r_3 = 3, $ and $r_4 = 3$, which corresponds to a five-petal shape.  Figure \ref{PR2Evo} shows the drop deformed in (a) shear flow and  (b) strain flow. For the shear flow,  $G=0.5$,  $\lambda=1.2$, $J_0=1$, $E=0.1$,  $K=1.5$, $\Gamma(\alpha,0)=0.6$,  $N_s=512$,  and $\Delta t=5 \times 10^{-4}$. Drop profiles are shown from  $t=0$ to $t=5.12$ with $\Delta t_p=0.512$. For the strain flow,  $Q=0.5$ with other parameters as in the shear flow except for
$K=1$ and $\Gamma(\alpha,0)=0.5$.  Drop profiles and surfactant concentration $\Gamma$ are shown at the same sequence of times.  
The number of interface points was doubled to $N_s=1024$ at time $t=1.28$.

\begin{figure}[h]
  
    \begin{subfigure}{.5\textwidth}
        \subcaption{}
        \centering
         \hspace{-.5in}
        \includegraphics[scale=.28]{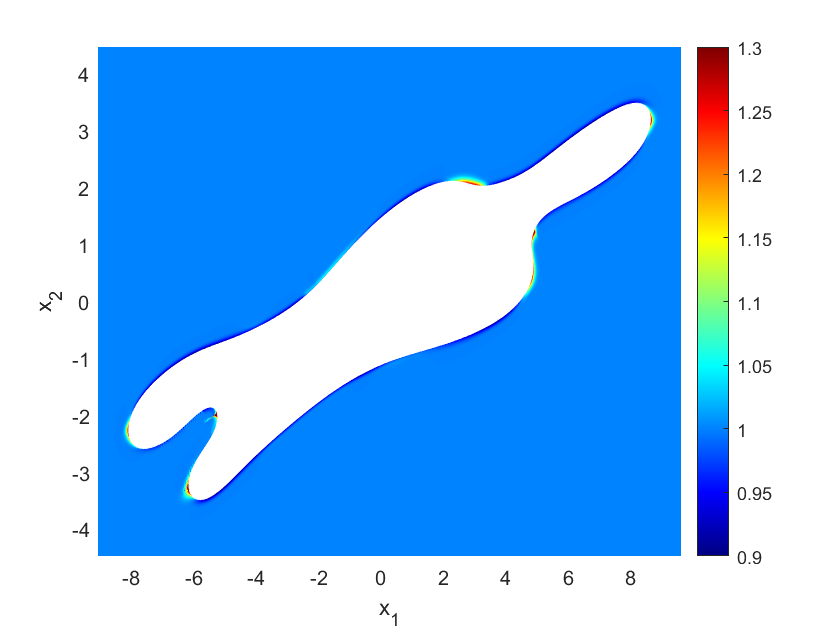}
        
        \label{pr2bulkshear}
    \end{subfigure}
    \hspace{-.4in}
    \begin{subfigure}{.5\textwidth}
        \subcaption{}
        \centering
         \hspace{-.5in}
        \includegraphics[scale=.39]{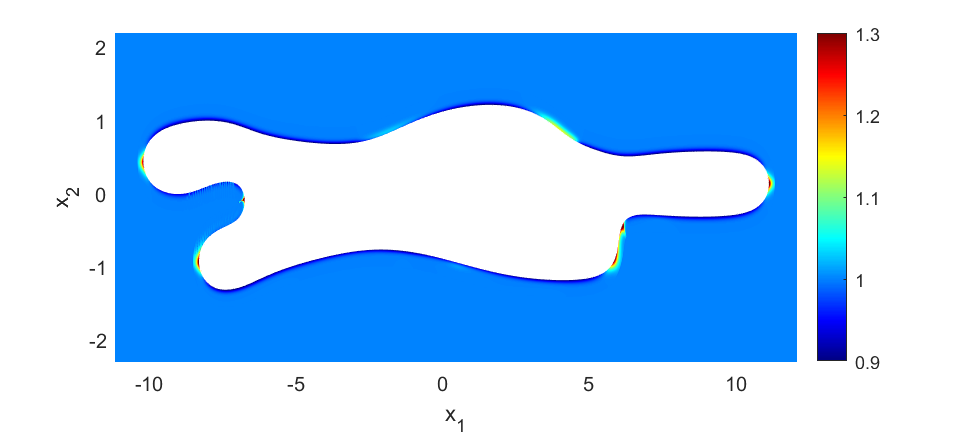}
        
        \label{pr2bulkstrain}
    \end{subfigure}
    \caption[Bulk concentration surfactant of a drop from Figure \ref{PR2Evo} at time $t=5.12$]{Bulk surfactant concentration for the simulations of	Figure \ref{PR2Evo} at $t=5.12$: (a) shear flow and (b) strain flow.}
    \label{PR2Bulk}
\end{figure}

Figure \ref{PR2Bulk} shows the bulk concentration of surfactant for the 
simulation of Figure \ref{PR2Evo} at $t = 5.12$. In the post-processing step, 
$M = 128$ with  
$\delta = 0.01$, and $Pe = 2000$. 
The highest bulk surfactant concentration 
occurs at regions of desorption and maximum interfacial surfactant concentration. 
The bulk concentration peaks at the three drop tips that persist
at this time. Elevated concentrations are also found near two tips that have relaxed due to contraction of the interface.

\subsection{C shape}

Next, we consider an initial C-shaped interface, previously studied in \cite{OJALA2015145} 
for the case of a clean (i.e., surfactant-free) drop.  The interface is parameterized by 
\begin{equation}\label{C dom para}
    \begin{aligned}
        x_1(\alpha) &= (b_1+\sin \alpha)\cos(b_2\pi\cos \alpha) \\
        x_2(\alpha) &= (b_1+\sin \alpha)\sin(b_2\pi\cos \alpha)
    \end{aligned}
\end{equation}
for $\alpha\in[0,2\pi)$.  In our simulations, we use the specific parameter values 
$b_1 = -1.5$ and $b_2 = 0.99$.

\begin{figure}[h]
\begin{subfigure}{.5\textwidth}
        \centering
        \includegraphics[scale=.32]{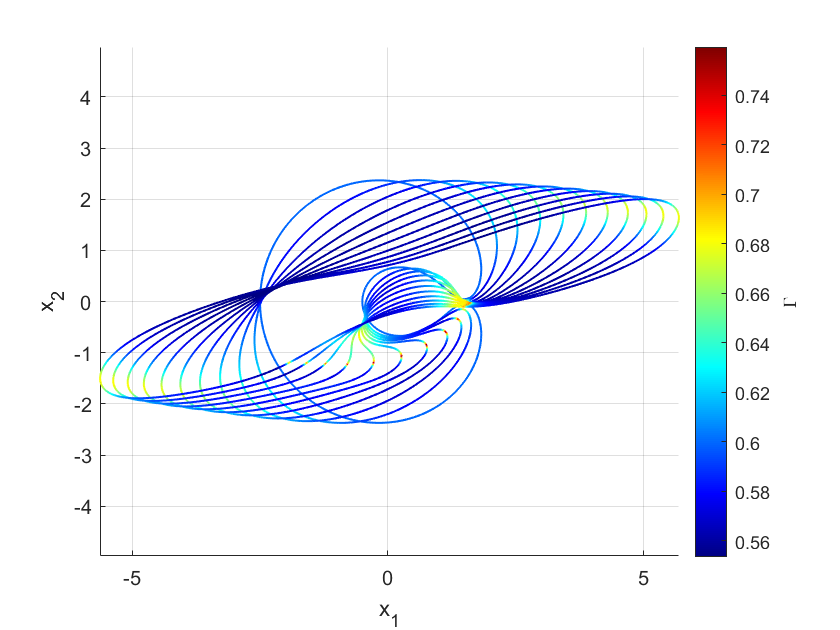}
        \put(-220,150){\bf (a)}
        
        \label{pr3shear}
    \end{subfigure}
    \begin{subfigure}{.5\textwidth}
        \centering
        \includegraphics[scale=.32]{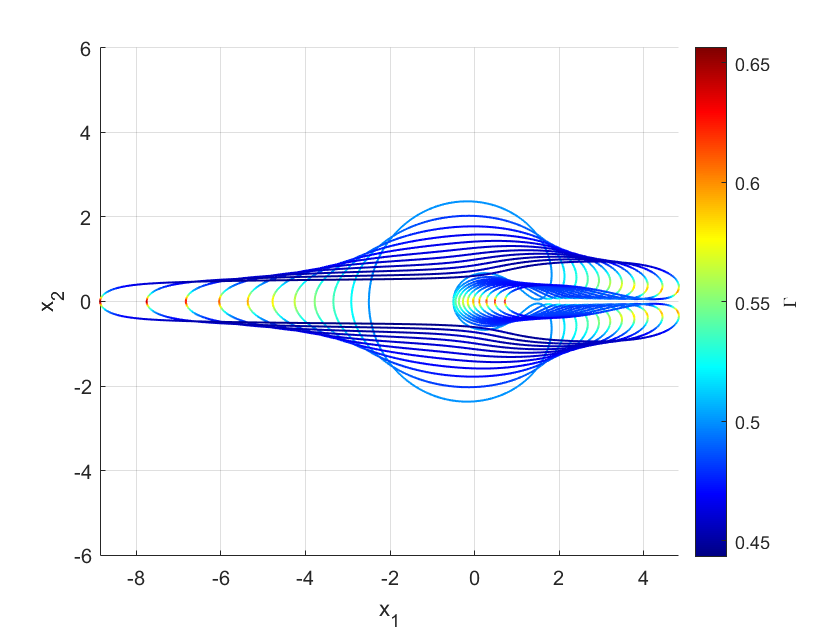}
         \put(-220,150){\bf (b)}
         
        \label{pr3strain}
    \end{subfigure}
    \caption[Deformation of drop from Figure \ref{CD Initial}]{Deformation of a C-shaped drop in (a) a shear flow,  and (b) a strain flow.}
    \label{CDevo}
\end{figure}

Figure \ref{CDevo} shows the deformation of the C-shaped drop in (a) a shear flow and (b) a strain flow. 
In both simulations
$\lambda = 1.2$, $E=0.1$, $J_0 = 1$, $N_s = 512$,  and $\Delta t = 5 \times 10^{-4}$. For the shear flow, $G = 0.5$, $K = 1.5$ 
with an initial surfactant distribution 
$\Gamma(\alpha, 0) = 0.6$,
while for  the strain flow  $Q = 0.5$, $K = 1$, and
$\Gamma(\alpha, 0) = 0.5$. To maintain accuracy, the number of interface points was increased 
to $N_s = 1024$ at $t = 1.92$. 

Drop profiles are shown from $t = 0$ to $t = 7.68$ at intervals of $\Delta t_p = 0.768$, with 
the interfacial surfactant concentration $\Gamma$ superimposed on each profile. 
In the strain flow, the drop interface is seen to approach self-intersection, resulting in near-singular 
integrals. Although specialized quadrature methods exist for accurately resolving such integrals \cite{paalsson2019simulation}, 
here we rely on adaptive interface refinement to ensure adequate resolution.

\begin{figure}[h]
    \begin{subfigure}{.5\textwidth}
        \subcaption{}
        \centering
        \includegraphics[scale=.32]{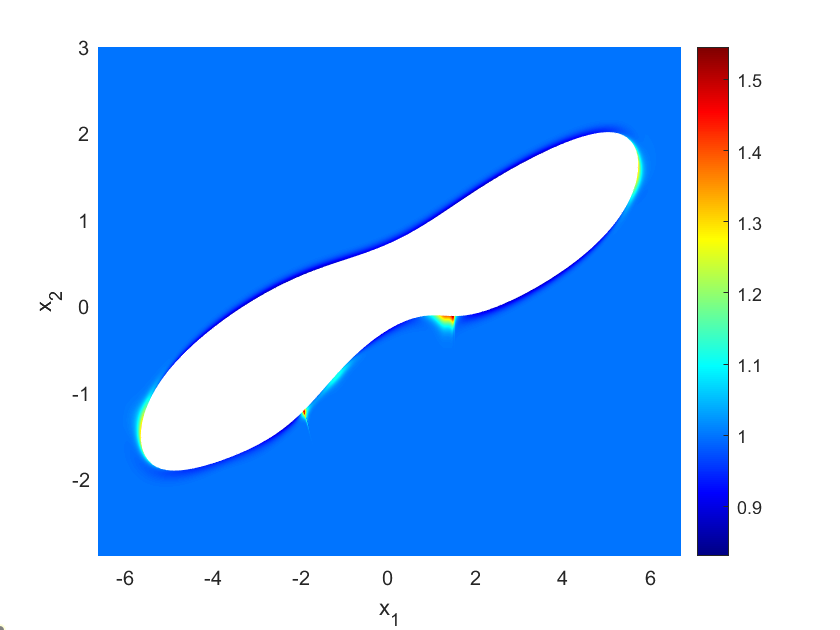}
        
        \label{cdbulkshear}
    \end{subfigure}
    \begin{subfigure}{.5\textwidth}
        \subcaption{}
        \centering
        \includegraphics[scale=.32]{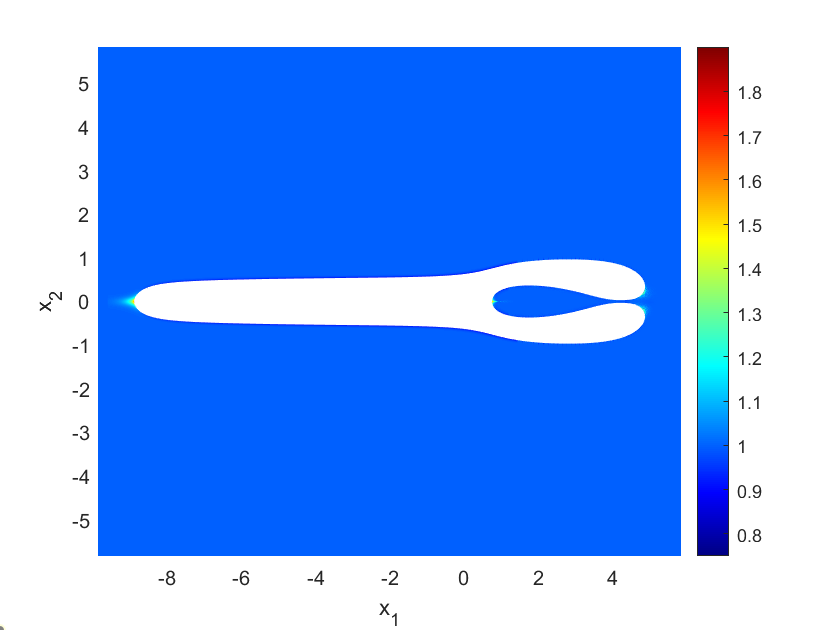}
        
        \label{cdbulkstrain}
    \end{subfigure}
    \caption[Bulk concentration of surfactant of a drop from Figure \ref{CD Initial} at time $t=7.68$]{Bulk concentration of surfactant for the simulations of Figure \ref{CDevo} at time $t=7.68$: (a) shear flow and (b) strain flow.}
    \label{CDBulk}
\end{figure}

Figure \ref{CDBulk} shows the bulk surfactant concentration at the final time $t = 7.68$. 
For the shear flow of Figure \ref{CDBulk}a, post-processing was performed with $M = 128$ points in the 
normal direction
and $\delta = 0.01$,  $Pe = 2000$. 
For the strain flow of Figure \ref{CDBulk}b, $M = 64$
with $\delta = 0.01$ and 
$Pe = 3000$.
As in previous examples, the highest bulk surfactant concentrations are observed near stagnation points of the flow on the interface, or at other regions where the interface undergoes significant contraction.

\subsection{Swiss-roll}\label{swiss}

As a third example, we consider an initial Swiss-roll shaped interface, which was introduced in \cite{OJALA2015145} and \cite{paalsson2019simulation} in the context of a clean and insoluble-surfactant--coated drop, respectively.
The initial drop shape is constructed from two one-dimensional spirals and two circular caps, with the entire interface smoothed via convolution with a heat kernel to ensure smoothness at the junctions between the caps and the spirals.

\captionsetup[subfigure]{font={bf,small}, skip=1pt, margin=-0.01cm, singlelinecheck=true}

The  simulation begins with $N_s = 1024$ interface points and a time step of $\Delta t = 3.9 \times 10^{-3}$. The flow parameters are set to $B = G = Q = 0$, so that no far-field flow is imposed and the drop evolves solely under the influence of surface tension. Additional parameters are $\lambda = 1$, $J_0 = 1$, $E=0.1$, $K = 1$,  with $\Gamma(\alpha,0) = 0.5$.
Since the drop shape relaxes to a circle, we can reduce the number of interface grid points without compromising accuracy. Accordingly, at time $t=25.3$, the number of grid points is halved to $N_s=512$. 

\begin{figure}[h!]
    \begin{subfigure}{.5\textwidth}
        \includegraphics[scale=.33]{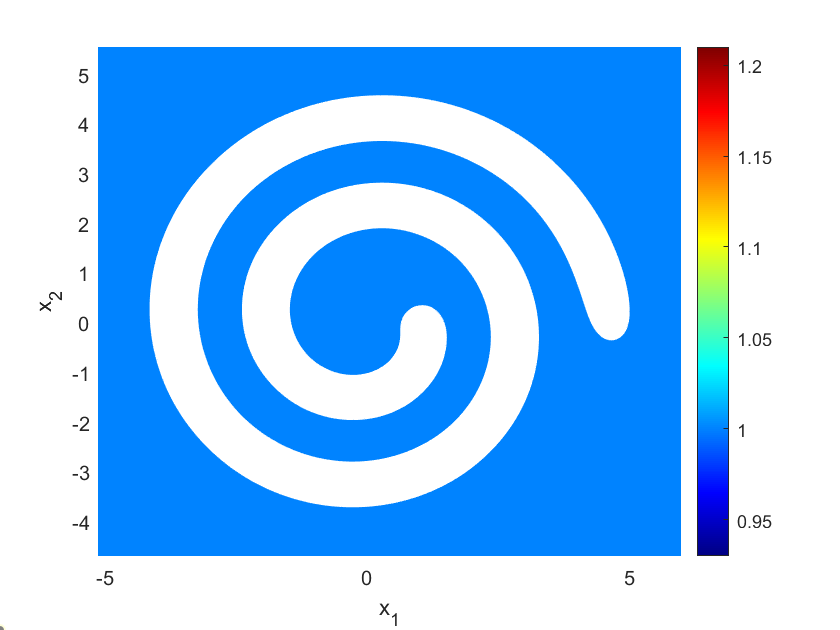}
        \caption{$t=0$}
    \end{subfigure}
   \begin{subfigure}{.5\textwidth}
        \includegraphics[scale=.33]{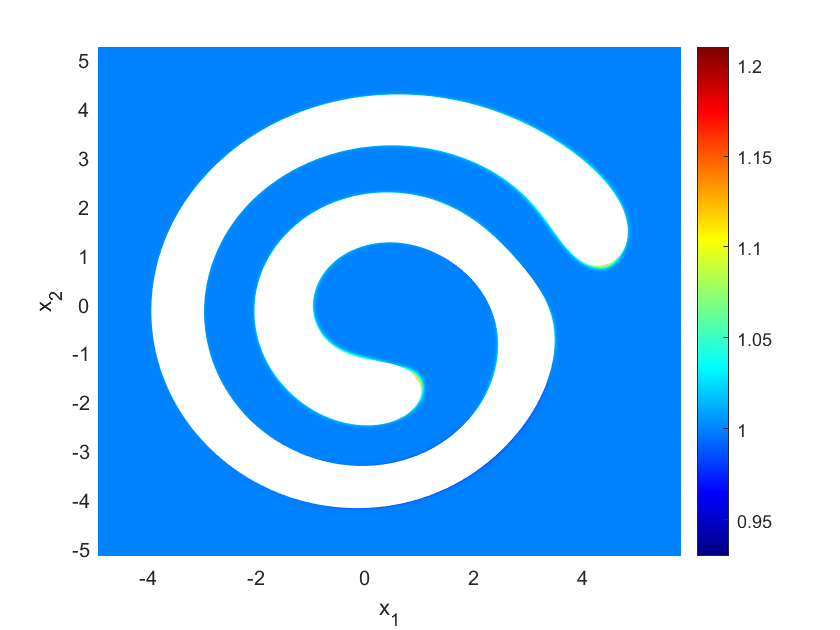}
        \caption{$t=5$}
    \end{subfigure}
    \begin{subfigure}{.5\textwidth}
        \includegraphics[scale=.33]{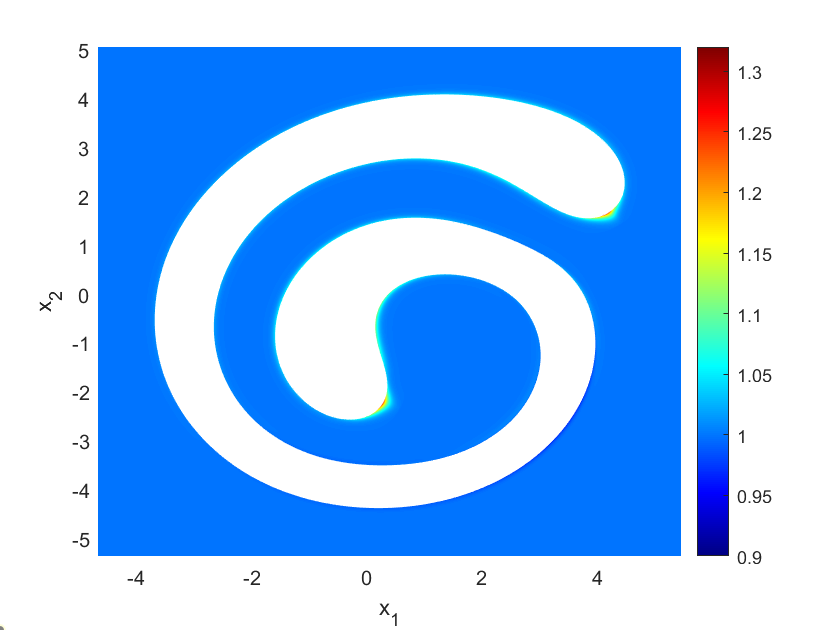}
        \caption{$t=10$}
    \end{subfigure}
   \begin{subfigure}{.5\textwidth}
        \includegraphics[scale=.33]{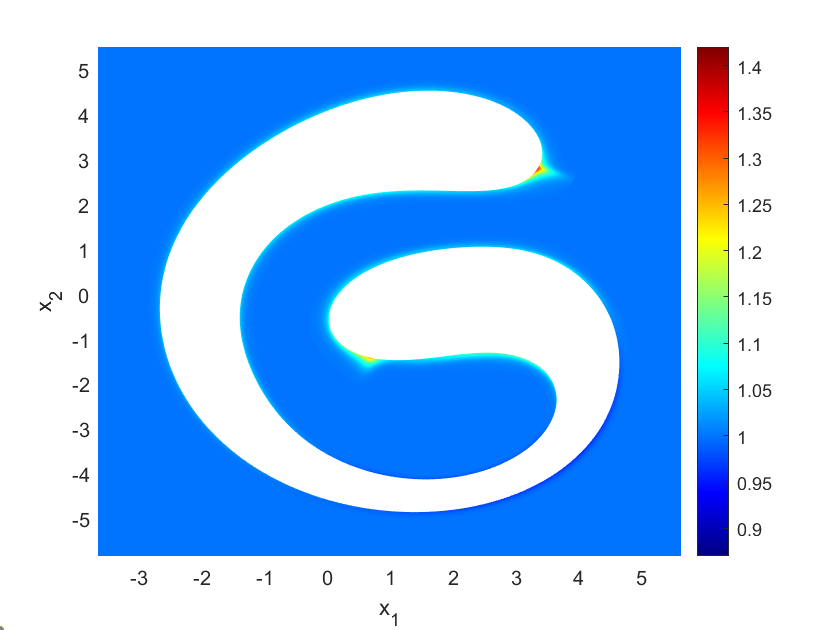}
        \caption{$t=20$}
    \end{subfigure}
    \begin{subfigure}{.5\textwidth}
        \includegraphics[scale=.33]{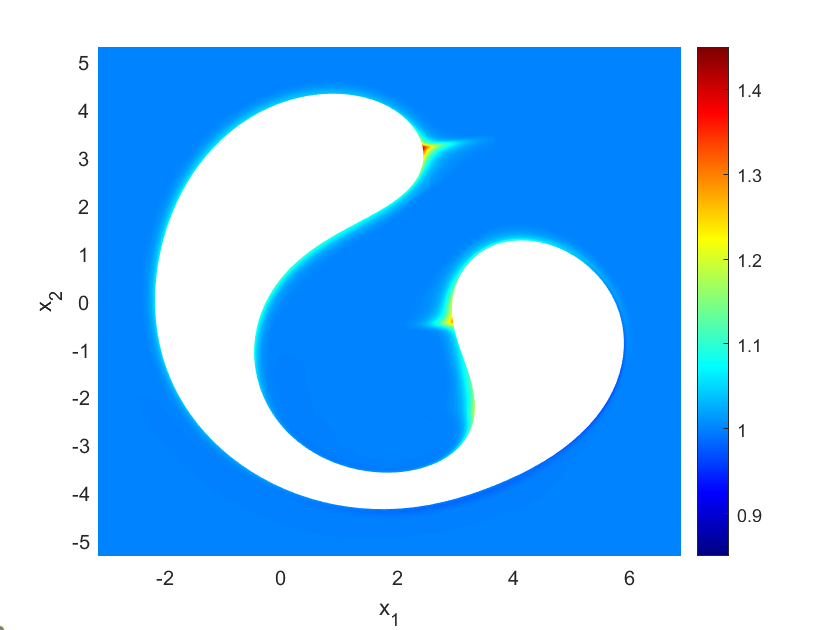}
        \caption{$t=30$}
    \end{subfigure}
   \begin{subfigure}{.5\textwidth}
        \includegraphics[scale=.33]{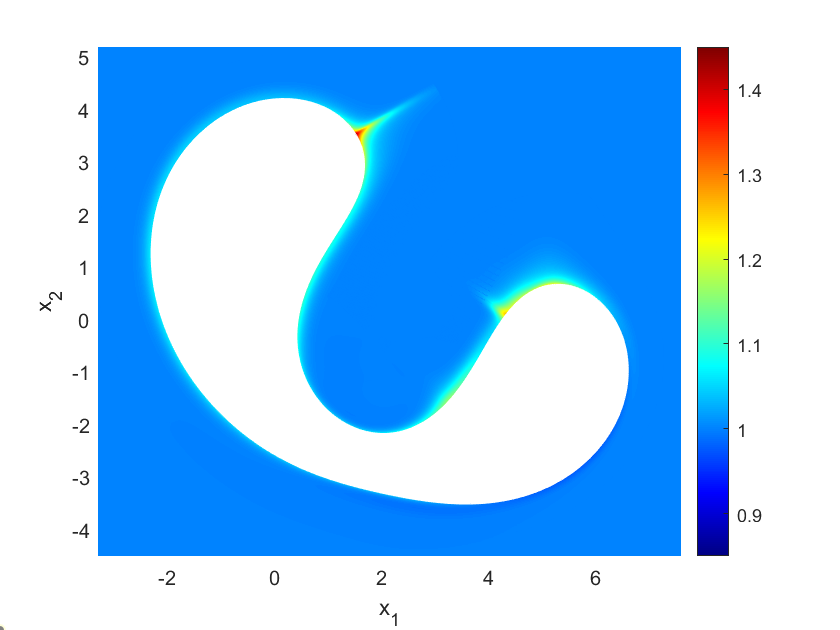}
        \caption{$t=40$}
    \end{subfigure}
    \begin{subfigure}{.5\textwidth}
        \includegraphics[scale=.33]{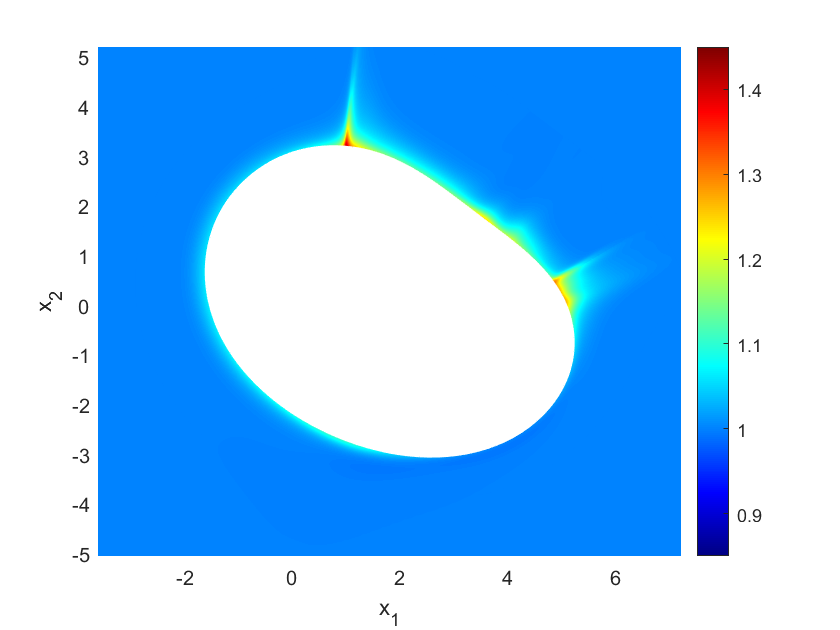}
        \caption{$t=60$}
    \end{subfigure}
    \begin{subfigure}{.5\textwidth}
        \includegraphics[scale=.35]{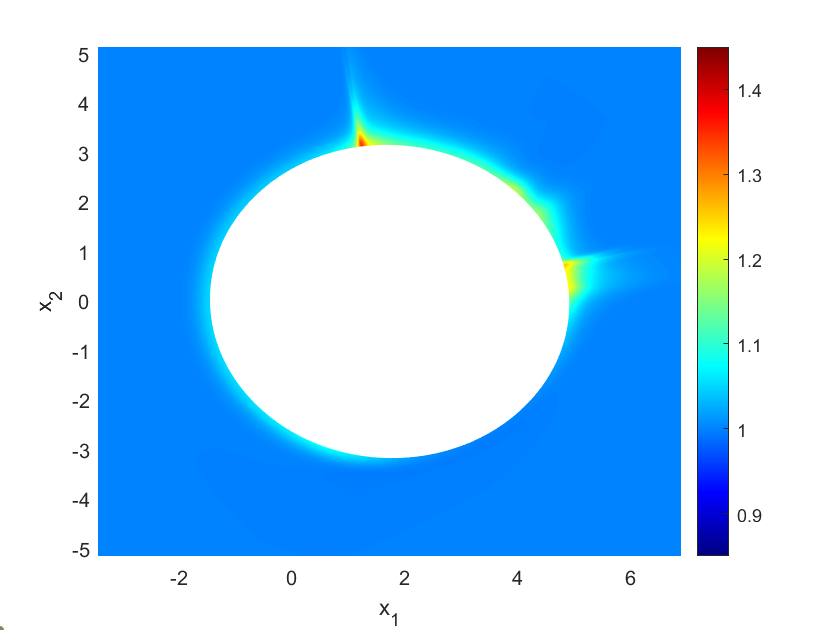}
        \caption{$t=80$}
    \end{subfigure}
    \caption[Swiss-roll drop relaxation with bulk surfactant concentration]{Relaxation of a Swiss-roll  interface under surface tension (first phase),  showing the evolution  of bulk surfactant concentration $C$.}
    \label{Sprial Bulk Relax}
\end{figure}

The results are illustrated in Figure \ref{Sprial Bulk Relax}, which presents snapshots of the drop evolution together with the bulk surfactant concentration $C$ at times $t = 0, ~5, ~10, ~20, ~30, ~40, ~60$, and $80$, with $Pe = 4000$. Post-processing is performed with $M = 256$ points in the normal direction.

 The dynamics progress through two distinct phases. In the first phase, which is shown in Figure \ref{Sprial Bulk Relax}, the capillary stress term $\sigma\kappa \mathbf{n}$ in the stress balance \eqref{stressbal} induces a significant normal velocity at the caps, causing the drop to contract. As seen in the sequence of snapshots, the interface gradually evolves to a nearly circular shape by time $t=80$ in panel (h).  Beyond this time, in the second phase ($t > 80$, not shown in Figure \ref{Sprial Bulk Relax}) the shape remains essentially unchanged, with the normal velocity of the interface $u_n \approx 0$ and the surface tension $\sigma$ nearly spatially uniform.

In this second phase the dynamics is primarily driven by the Marangoni stress term $-\nabla_s \sigma$ on the right-hand side of \eqref{stressbal}. This induces a spatially varying tangential flow along the nearly circular interface, promoting the redistribution of surface surfactant concentration $\Gamma$ toward a uniform state. As a result, $\Gamma$ gradually becomes spatially uniform over time.

\begin{figure}[h]
    \centering
    \includegraphics[scale = 0.4]{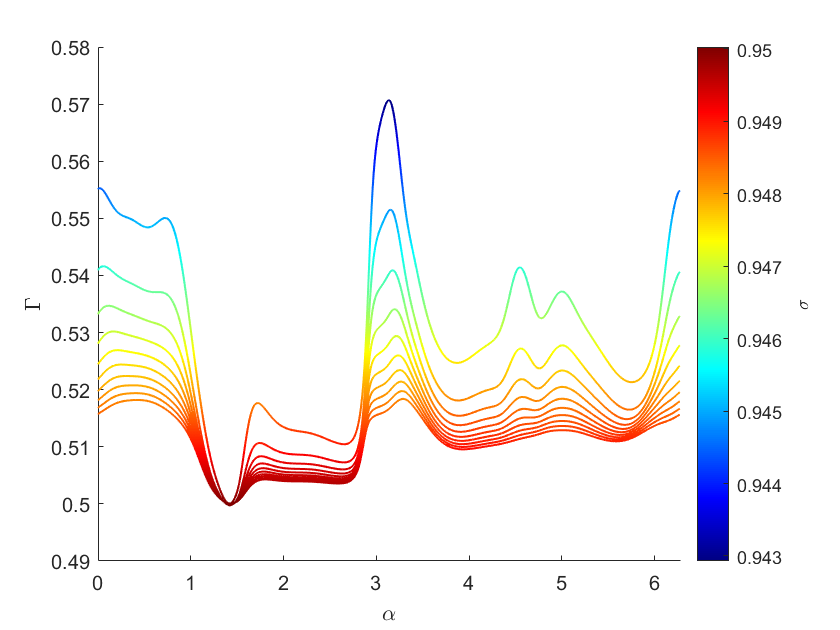}
    \caption[Surface surfactant concentration $\Gamma$ with surface tension $\sigma$ superimposed at long times]{Surface surfactant concentration $\Gamma$ versus $\alpha$  from time $t=81.92$ to $t=819.2$ (second phase) with $\Delta t_p=81.92$ and surface tension $\sigma$ superimposed.  There is a slow decrease of $\Gamma$ toward an equilibrium value  during this phase.}
    \label{Spiral Gamma}
\end{figure}

Figure \ref{Spiral Gamma} illustrates this second phase evolution of $\Gamma$ versus $\alpha$, from $t = 81.92$ to $t = 819.2$ at intervals of $\Delta t_p = 81.92$, with the surface tension $\sigma$ superimposed. This shows a slow monotone decrease in the spatial variation of $\Gamma$, driven by Marangoni-induced advection, so that $\sigma$ becomes more spatially uniform and increases in magnitude. During this phase, surface diffusion of $\Gamma$ is relatively weak; instead, advection of $\Gamma$ along the interface, together with advection and diffusion of the bulk surfactant concentration $C$ near the interface reduce the spatial gradient of $\Gamma$.

From the equation of state \eqref{eqstate}, the Marangoni stress  $-\nabla_s \sigma = E \nabla_s \Gamma$, so that the time scale for approach to a uniform distribution of  $\Gamma$ is proportional to $1/E \simeq 10$ which is significantly longer than the time scale of the first phase.

\clearpage

\section{Conclusion} \label{sec:Conclusion}

We have developed a fast mesh-free method for two-phase flow with soluble surfactant at large bulk P\'eclet number. This is based on a new method for fast computation of an Abel-type time-convolution integral for the bulk surfactant concentration, and a boundary integral formulation of \cite{xu2013analytical}.  The algorithm overcomes several difficulties in the efficient computation of the time convolution, including the dependence of the kernel on the time history of the evolution.  

An important aspect of the algorithm is its provision of a fast surface-based formulation for the full interfacial flow problem, 
including advection and diffusion of soluble surfactant in the large Péclet number limit, which is itself of intrinsic interest. The method for computing the time-convolution integral has complexity  $O(P\log_2^2 P)$ per interface point,  where $P$ is the number of time steps, compared with $O(P^2)$ for direct quadrature. The order of accuracy of the method as implemented is $O(h^{3/2})$, where $h$ is the time-step size, but in principle this can be increased by incorporating higher order expansions in  \eqref{discretize0} and (\ref{first_time_step}). The method is validated by comparison with a previously developed mesh-based approach of \cite{xu2013analytical}. 
Compared to the mesh-based approach, the fast mesh-free method avoids the use of a grid in the direction normal to the interface and eliminates the need for artificial truncation of the transition layer domain.   However, this comes at the cost of a more intricate implementation.

To simplify the implementation of the mesh-free method several assumptions have been made.  Among these,  the formulation is restricted to a single drop in a two-dimensional geometry, with the soluble surfactant present solely in the exterior fluid. Outside the transition layer the bulk surfactant concentration is considered uniform. At the interface we apply the boundary condition (\ref{BCbulkC}), which corresponds to the diffusion-controlled kinetics limit of the more general mixed-kinetic expression. Nevertheless, the fast algorithm for computing the time-convolution integral can be extended with only minor modification to a three-dimensional geometry and multiple interacting drops. Spatially varying surfactant concentrations, either outside the boundary layer or in the drop interior,  can be treated as in  \cite{atwater2020studies} or by a semi-Lagrangian approach.  In principle, the method can also be extended to the more general mixed-kinetic boundary condition, see for example \cite{wang2014numerical}.

Applications of the method described here extend beyond the context of interfacial flow with surfactant to other problems that feature advection-diffusion of a bulk quantity with small diffusion, where surface activity of the quantity affects the flow dynamics and evolution of the interface.  This occurs, for example, during
diffuseophoresis of colloidal particles \cite{yariv2015phoretic} and in mass and charge transport within electrochemical cells \cite{braff2013boundary}.   Similar Abel-type time convolutions also arise in different contexts, including the Green's function formulation of the Dyson equation from quantum many-body physics \cite{kaye2021low}.   The methods developed here have the potential to impact these and other areas.  

\section{Acknowledgements} 
We are grateful to the National Science Foundation for support through grant DMS-1909407 (SE, MS and MB). MS was also supported by the Petroleum Research Fund of the American Chemical Society through grant PRF\#68292-ND9, and by an NJIT Seed Grant. 

\begin{appendices}
\section*{Appendix A ~~Validation of the fast time convolution using synthetic data }\label{APP Synthetic} 
As an additional check on the accuracy of the fast algorithm described in Section \ref{FastHM}, we apply the method to a synthetic example that has an exact analytical representation for the time-convolution integral of \eqref{kernel}, i.e.,
\begin{equation}\label{kernelA}
    \mathcal{K} g(t) = \int_0^t \frac{1}{\sqrt{t-\tau}} k(t,\tau)  g(\tau) \, d\tau,
\end{equation}
where $k(t,\tau)$ is given by (\ref{k(t,tau)}).

 Set
\begin{equation}\label{synpsi0}
    \psi_0(t) = 1-\frac{1}{2t+2},
\end{equation}
so that the solution of \eqref{psiODE} is
\begin{equation}\label{synpsi1psi2}
    \begin{aligned}
        \psi_1(t) = t-\half\log[1+t]\\
        \psi_2(t) = \frac{3}{4}-\frac{3+2t}{4}\exp[-2t].
    \end{aligned}
\end{equation}
Next,  choose $g(\tau)=g_\alpha(\tau)=
  \psi_2'(\tau)\psi_2^\alpha(\tau)$ and substitute $x = \psi_2(\tau)/\psi_2(t)$ into (\ref{kernelA}) to obtain
\begin{equation}\label{synanalytic}
\begin{aligned}
    \frac{1}{\sqrt{\pi}}\exp[-\psi_1(t)]\int_{0}^{t}\frac{\psi_2'(\tau)\psi_2^\alpha(\tau)}{\sqrt{\psi_2(t)-\psi_2(\tau)}} d\tau &= \frac{1}{\sqrt{\pi}}\exp[-\psi_1(t)]\int_{0}^{1}\frac{\psi_2'(t)x^\alpha \psi_2^\alpha(t)\psi_2(t)}{\psi_2'(t)\sqrt{\psi_2(t)-x\psi_2(t)}}  \ dx, \\
    &= \frac{1}{\sqrt{\pi}}\exp[-\psi_1(t)]\psi_2^{\alpha+1/2}(t)B(\half,1+\alpha),
\end{aligned}
\end{equation}
where $B$ is the Euler beta function.

To test the convergence of the quadrature method, we consider values of $\alpha$  that  lead to the following asymptotic expansions of $g_\alpha(\tau)$ near $\tau=0$:

\begin{align*}
\alpha= -\half\quad &\Rightarrow\quad g_{-\half}(\tau) = \frac{1}{\sqrt{\tau}} -
                 \frac{3}{4}\sqrt{\tau} + O(\tau^{3/2}),\\
\alpha= 0\quad &\Rightarrow\quad g_{0}(\tau) = 1 - \tau + O(\tau^3), \\
\alpha= \half\quad &\Rightarrow\quad g_{\half}(\tau) = \sqrt{\tau} + O(\tau^{3/2}),\\
\alpha= 1 \quad &\Rightarrow\quad g_{1}(\tau) = \tau + O(\tau^2), \\
\alpha= \frac{3}{2}  \quad &\Rightarrow\quad g_{\frac{3}{2}}(\tau) = \tau^{\frac{3}{2}} + O(\tau^\frac{5}{2}).
\end{align*}
Based on these expansions, we consider the following examples
\begin{align}
  E0:\quad  g(\tau) &= g_{-\half}(\tau) + g_0(\tau), \nonumber \\
  E1:\quad  g(\tau) &= g_0(\tau), \nonumber \\
  E2:\quad  g(\tau) &= g_{\half}(\tau), \label{synalpha}\\
  E3:\quad  g(\tau) &= g_{1}(\tau), \nonumber \\
  E4:\quad  g(\tau) &= g_{\threehalf}(\tau). \nonumber
\end{align}
In all examples, we  use  $\psi_0,~ \psi_1$ and $\psi_2$ as given in (\ref{synpsi0}) and   (\ref{synpsi1psi2}).
The numerical results obtained are shown in
Tables~\ref{tab:fast} and~\ref{tab:slow}. It is seen  that the
complexity of the fast method is consistent with log-linear behavior in $P$,
while the direct method is quadratic. The interpolation order $q$ of the fast 
method in \eqref{kernel:OneVar} is chosen such that the errors of the
fast and direct method do not differ in the first three significant digits.

Multiplication of (\ref{synalpha}) by (\ref{varphi2}) shows that the integrands of examples E0, E1 and E3 have a nonzero  linear or  $O(\tau)$ term, 
while the linear term of E2 and E4 vanishes. As seen in
Tables~\ref{tab:fast} and~\ref{tab:slow}, the   former  produce an
$O(h^{3/2})$ discretization error whereas the latter produce an
$O(h^2)$ error, which matches the expected accuracy.

\begin{table}[hbt]
\begin{center}
\begin{tabular}{|c|ccccccc|}
\hline
P & q & operations & E0 & E1 & E2 & E3 & E4    \\
\hline
  40  & 2 & 5.49e+02  & 2.27e-02 & 2.09e-02 & 6.68e-03 & 2.59e-02 & 6.14e-03\\
   80 & 3 & 1.84e+03  & 7.75e-03 & 7.33e-03 & 1.66e-03 & 9.41e-03 & 1.60e-03\\
  160 & 3 & 5.12e+03  & 2.68e-03 & 2.58e-03 & 4.15e-04 & 3.38e-03 & 4.06e-04\\
  320 & 4 & 1.55e+04  & 9.36e-04 & 9.10e-04 & 1.04e-04 & 1.20e-03 & 1.03e-04\\
  640 & 4 & 3.97e+04  & 3.28e-04 & 3.21e-04 & 2.59e-05 & 4.26e-04 & 2.58e-05\\
 1280 & 5 & 1.10e+05  & 1.15e-04 & 1.14e-04 & 6.47e-06 & 1.51e-04 & 6.46e-06\\
 2560 & 5 & 2.66e+05  & 4.05e-05 & 4.01e-05 & 1.62e-06 & 5.34e-05 & 1.62e-06\\
 5120 & 6 & 6.94e+05  & 1.43e-05 & 1.42e-05 & 4.04e-07 & 1.89e-05 & 4.04e-07\\
  \hline
&&$O(P \log_2^2 P)$ & $O(h^{3/2})$& $O(h^{3/2})$& $O(h^2)$&  $ O(h^{3/2})$& $O(h^2)$\\
  \hline
\end{tabular}
\caption{Errors and complexity for the  test cases with the  fast time-convolution method}
\label{tab:fast}
\end{center}
\end{table}

\begin{table}[hbt]
\begin{center}
\begin{tabular}{|c|cccccc|}
\hline
P & operations & E0 & E1 & E2 & E3 & E4    \\
\hline
  40  & 8.19e+02 &   2.27e-02 & 2.09e-02 & 6.68e-03 & 2.59e-02 & 6.14e-03\\
   80 & 3.24e+03 &   7.75e-03 & 7.33e-03 & 1.66e-03 & 9.41e-03 & 1.60e-03\\
  160 & 1.29e+04 &   2.68e-03 & 2.58e-03 & 4.15e-04 & 3.38e-03 & 4.06e-04\\
  320 & 5.14e+04 &   9.36e-04 & 9.10e-04 & 1.04e-04 & 1.20e-03 & 1.03e-04\\
  640 & 2.05e+05 &   3.28e-04 & 3.21e-04 & 2.59e-05 & 4.26e-04 & 2.58e-05\\
 1280 & 8.20e+05 &   1.15e-04 & 1.14e-04 & 6.47e-06 & 1.51e-04 & 6.46e-06\\
 2560 & 3.28e+06 &   4.05e-05 & 4.01e-05 & 1.62e-06 & 5.34e-05 & 1.62e-06\\
 5120 & 1.31e+07 &   1.43e-05 & 1.42e-05 & 4.04e-07 & 1.89e-05 & 4.04e-07\\
  \hline
&$O(P^2)$ &  $O(h^{3/2})$& $O(h^{3/2})$& $O(h^2)$&  $ O(h^{3/2})$& $O(h^2)$\\
  \hline
\end{tabular}
\caption{Errors and complexity for the test cases with the direct time-convolution method}
\label{tab:slow}
\end{center}
\end{table} 

\end{appendices}

 \bibliographystyle{elsarticle-num} 
\newpage 
\bibliography{refscombined.bib}

\end{document}